\documentclass[12pt]{article}
\usepackage{amsmath}
\usepackage{amsthm}

\usepackage{amssymb}
\usepackage{wasysym}
\usepackage{graphicx}
\usepackage{comment}
\usepackage{color}
\usepackage{multicol}
\newtheorem{lemma}{Lemma} 
\newtheorem{remark}{Remark}
\newtheorem{theorem}{Theorem}
\DeclareMathOperator*{\argmin}{arg\,min}
\begin{document}
%
\title{Sparse constrained projection approximation subspace tracking}
%
%
%

\author{Denis~Belomestny, 
        Ekaterina~Krymova
\thanks{The authors are with Faculty of Mathematics at the University  Duisburg-Essen, Essen, Germany}
\thanks{D.\,Belomestny e-mail: denis.belomestny@uni-due.de}
\thanks{E.\,Krymova e-mail: ekaterina.krymova@uni-due.de}
\thanks{This work is funded by the German Research Foundation (DFG), Collaborative Research Center 823, Subprojects B3 and C5. The first author was also supported  by the RSF grant 18-11-00132}
}
\maketitle

\begin{abstract}
In this paper we revisit the well-known constrained projection approximation subspace tracking algorithm (CPAST) and derive, for the first time, non-asymptotic error bounds. Furthermore, we introduce a novel sparse modification of CPAST   which is able to  exploit sparsity in the underlying covariance structure. We present a non-asymptotic analysis of  the proposed algorithm and study its empirical performance on simulated and real data. 
\end{abstract}


%

\section{Introduction}
Subspace tracking methods are intensively used in statistical and signal processing community.
Given  observations of a multidimensional signal, one is interested in  estimating or tracking a subspace spanning the eigenvectors corresponding to the first largest eigenvalues of the signal covariance matrix. 
Over the past few decades many variations  of the original projection approximation subspace tracking (PAST) method   \cite{yang1995projection}  
were developed which found applications in data compression, filtering,
speech enhancement, etc.  (see \cite{delmas2010subspace} and references therein).
 Despite  popularity of  the subspace tracking  methods, only partial results are
known about their convergence. 
The asymptotic convergence of the PAST algorithm was first established in \cite{yang1996convergence,yang1996asymptotic} using a general theory of stability for ordinary differential equations. However,  no finite sample error bounds are available in the literature.
Furthermore, in the case of a high-dimensional signal  the empirical covariance matrix estimator  performs poorly if  the number of observations is small. A common way to improve the estimation quality in this case is to impose  some kind of  sparsity assumptions on the signal itself or on the eigensubspace of the underlying covariance matrix. 
In  \cite{ma2013sparse} a sparse modification of the orthogonal iteration scheme for a fixed number of observations was proposed. A thorough analysis in \cite{ma2013sparse} shows that under appropriate  sparsity assumptions on the leading eigenvectors, the orthogonal iteration scheme combined with thresholding  allows to perform dimension reduction in high-dimensional setting. Our main goal is to propose a novel modification of  constraint projection approximation subspace tracking method (CPAST) \cite{valizadeh2009fast}, called sparse projection approximation subspace tracking method (SCPAST),  which can be used for efficient subspace tracking  in the case of high-dimensional sparse signal and small number of available observations. Another contribution of our paper is a non-asymptotic convergence analysis of CPAST and SCPAST algorithms showing the advantage of SCPAST algorithm in the case of sparse covariance structure. Last but not the least, we analyse  numerical performance of SCPAST algorithm on simulated and real data. In particular, the problem of tracking the leading subspace of a music signal is considered. 
\par
The structure of the paper is as follows. In Section~\ref{sec:main_setup} we introduce our observational model and formulate main assumptions. Section~\ref{sec:cpast} first reviews the CPAST algorithm   and then provides the non-asymptotic error bounds for  CPAST  in a "stationary" case. In Section~\ref{sec: SCPA}  we introduce our sparse constraint approximation subspace tracking method  and prove the non-asymptotic upper bounds for the estimation error.  A  numerical study of the proposed algorithm is presented in Section~\ref{sec:sim}.  Finally the proofs are collected in Section~\ref{sec:proofs}.

\section{Main setup}  
\label{sec:main_setup}
One important  problem in signal processing  is  adaptive estimation of
a dominant subspace given  incoming noisy observations.  
Specifically one considers a model
\begin{equation}
\label{init_model}
x(t)=s(t)+\sigma(t)\xi(t),\quad t=1,\dots,T,
\end{equation}
where the observations $x(t)\in\mathbb{R}^{n}$  contain the signal  $s(t)\in\mathbb{R}^{n}$ corrupted by a  vector $\xi(t)\in \mathbb{R}^n$  with independent
standard Gaussian components. The signal $s(t)$ is usually modelled as
\[
s(t)=A(t)\eta(t),
\]
where $A(t)$ is a deterministic
$n\times d$ matrix  of rank $d$ with  $d\ll n$
and $\eta(t)$
is a random vector  in $\mathbb{R}^d$   independent of \(\xi(t),\)  
such that $\mathbf{E}[\eta(t)]=0$ and $\mathbf{E}[\eta_{i}^{2}(t)]=1$, $i=1,\dots,d$. 
Under these assumptions, the process  $x(t)$ has a covariance  matrix $\Sigma(t)$ which may be decomposed in the following way
\begin{equation}
\label{covariance_obs}
\Sigma(t)=\mathbf{E}[x(t)x^{\top}(t)]=A(t)A^{\top}(t)+\sigma^{2}(t)I_{n},
\end{equation}
where \(I_{n}\) stands for  the unit matrix in \(\mathbb{R}^n.\)
Note that the matrix $ A(t)A^{\top}(t)$ has the rank $d$ and by the
singular value decomposition (SVD)
\begin{align*}
 A(t)A^{\top}(t) & =\sum_{i=1}^{d}\lambda_{i}(t)v_{i}(t)v_{i}^{\top}(t),
\end{align*}
where  $ v_i(t)\in\mathbb{R}^n, \,i=1,\dots,d,$ are the eigenvectors of $A(t)A^{\top}(t)$ corresponding to the eigenvalues $\lambda_{1}(t)\geq\lambda_{2}(t)\geq\dots\geq\lambda_{d}(t)>0$. 
It follows from \eqref{covariance_obs} that the first $d$ eigenvalues of $\Sigma(t)$ are 
$
\lambda_1(t) + \sigma^2(t), \dots,\lambda_d(t) + \sigma^2(t),
$ 
whereas the remaining $n-d$ eigenvalues are equal   to $\sigma^2(t)$. 
Since \mbox{$\lambda_{d}(t)>0,$} the subspace corresponding to the first $d$ eigenvectors of $ A(t)A^{\top}(t)$ is identifiable. 
\textit{
The subspace tracking methods aim to estimate the subspace  $\mathrm{span}(v_{1}(t),\ldots, v_d(t))$  based on the  observations $(x(k))_{k=1}^t.$ The overall number of observations $T$ is assumed to be fixed and known.}
\par
Relying on a  heuristic   assumption of slow (in time) varying 
 $\Sigma(t)$, the subspace tracking methods use the following 
estimator of the  covariance matrix (up to scaling)   
\begin{equation}
\label{c_gamma}
 \widehat{\Sigma}_{\gamma}(t)=\sum_{i=0}^{t}\gamma^{t-i}x(i)x^{\top}(i),
\end{equation}
where $0<\gamma\leq 1$ is the so-called forgetting factor. 
The estimator  $ \widehat{\Sigma}_{\gamma}(t)$ can adapt to the change in $\Sigma(t)$  
by discounting the past observations. 
In the stationary regime, that is,  if  $\Sigma(t)$ is a constant matrix, one would use $\gamma=1$.
It is well known, that in the case of Gaussian
independent noise  the estimator $\widehat{\Sigma}_{1}(t)$ 
is consistent.

%
%
%
%

 

\section{CPAST} 
\label{sec:cpast}

For the general model \eqref{init_model} and non-stationary case, constrained projection approximation subspace tracking (CPAST) method allows to iteratively  compute   a matrix $\widehat{V}_{\gamma}(t)$, $t=1,\dots,T$, containing the first $d$ leading eigenvectors  of the matrix $\widehat{\Sigma}_{\gamma}(t)$ (see  \eqref{c_gamma})  based on  sequentially arriving  observations $x(j)$, $j=1,\dots,t.$  The procedure starts with some initial approximation $\widehat{V}_{\gamma}(0)=\widehat{V}^0$ and consists of the following   two steps   
\begin{itemize} 
\item \textit{multiplication}: compute the $n\times d$ matrix 
\[
\widehat{\Sigma}_{\gamma,V}(t)=\widehat{\Sigma}_{\gamma}(t)\widehat{V}_{\gamma}(t-1);
\]
\item \textit{orthogonalization}: compute an estimator $\widehat{V}_{\gamma}(t)$ of the matrix $V(t)$ containing $d$ leading eigenvectors via 
\[
\widehat{V}_{\gamma}(t)=\widehat{\Sigma}_{\gamma,V}(t)[\widehat{\Sigma}_{\gamma,V}^{\top}(t)\widehat{\Sigma}_{\gamma,V}(t)]^{-1/2}.
\]
\end{itemize} 

In the "stationary" case ($\gamma=1$) the method  may be regarded as the "online"-version of the orthogonal iterations scheme (see \cite{golub2012matrix}) for computing the eigen-subspace of the non-negatively definite matrix.  With the use of the Sherman-Morrison-Woodbury formula for the inversion at each time $t,$ one has to perform 
$O(nd^2)$ operations to compute the updated matrix $\widehat{V}_{\gamma}(t)$ given $\widehat{V}_{\gamma}(t-1)$, $\widehat{\Sigma}_{\gamma}(t-1)$ and $x(t)$. 
\subsection{Convergence of CPAST}

Throughout this section we consider the stationary case where $\Sigma(t) = \Sigma$, $A(t)=A$, $v_i (t)=v_i$, $\lambda_i(t)=\lambda_i$, $i=1,\dots,d$, $\sigma^2(t)=\sigma^2$. In this situation one would like to keep all the available information to estimate $V$, that is,  to use the estimator \eqref{c_gamma} for $\Sigma$ with $\gamma=1$.  For notational simplicity, from now on  we skip the dependence on $\gamma$ and use the notation  $\widehat{\Sigma}(t)$ for the  empirical covariance matrix \(\widehat{\Sigma}(t)=\frac{1}{t}\sum_{i=1}^{t} x(i)x^{\top}(i)\) and $\widehat{V}(t)$ for  CPAST estimator. Thus in  the stationary case   the CPAST estimator takes the form
\begin{equation}
\label{cpast_1}
\widehat{V}(t) = [\widehat{\Sigma}(t) \widehat{V}(t-1)][\widehat{V}^{\top}(t-1)^{\top}\widehat{\Sigma}^2(t)\widehat{V}(t-1)]^{-1/2}.
\end{equation}
We assume that  the random vectors $\eta(t)$ and $\xi(t)$ have independent $\mathcal{N}(0,1)$ components for $t=1,\dots,T$.  Under these assumptions the covariance matrix \eqref{covariance_obs} becomes 
\begin{equation}
\label{stationary_C}
\begin{split}
\Sigma 
 =\sum_{i=1}^d\lambda_i v_i v_i^{\top}+\sigma^2 I_n=V \Lambda_d V^{\top}+\sigma^2 I_n, 
 \end{split}
\end{equation}
where $V$ is  $n\times d$ matrix with columns $\{v_i\}_{i=1}^d$, $\Lambda_d$ is  $d\times d $  diagonal   matrix with $\{\lambda_i\}_{i=1}^{d}$ on the diagonal.
Note that the observational model \eqref{init_model} in stationary case can be alternatively written as the so-called spike model  
\begin{equation}
\label{stationary_model}
x(t)=\sum_{i=1}^d\sqrt{\lambda_{i}}u_i(t) v_i+ \sigma \xi(t),  
\end{equation}
where $u_i(t)$ are i.i.d. standard Gaussian random variables independent from $\xi(t)$.
\par
For our non-asymptotic error analysis of CPAST, we assume that $d$, $\sigma^2$ and $\lambda_{i}$, $i=1,\dots,d$  are known. 
With the known $\sigma^2$ we can always normalize the data and therefore without loss of generality we can assume that $\sigma^2=1$. 
\par
The typical condition while analyzing the quality of the eigenvectors estimation is the so-called spectral gap condition, which says that the adjacent eigenvectors explain distinguishably different portion of the variance in the data, namely there exists $\tau\geq 1$, such that for all $j=1,\dots,d,$
$$
\tau (\lambda_j-\lambda_{j+1})\geq \lambda_1,
$$
where $\lambda_{d+1}=0$ by definition.
Since our goal is the estimation of the $d$-dimensional subspace of the first eigenvectors, and we are not interested in the estimation of each particular eigenvector, we need only the condition for the separation of this $d$-dimensional subspace, namely that the gap between $\lambda_d$ and $\lambda_{d+1}$ is sufficiently large:
\begin{equation}
\label{AD}
\tau  \lambda_d  \geq \lambda_1.
\end{equation}
Define a distance \(l\) between two subspaces $\mathcal{W}$ and $\mathcal{Q}$  spanning  orthonormal columns $ w_{1},\ldots,w_d$ and $ q_{1},\ldots, q_d $  correspondingly via 
\begin{equation}
\label{distance}
l({W}, {Q})=l(\mathcal{W},\mathcal{Q})=\|WW^{\top}-QQ^{\top}\|^{2},
\end{equation}
where the nuclear norm \(\|A\|\) of a matrix $A\in\mathbb{R}^{n\times d}$ is defined as \mbox{$\|A\|=\sup_{x\in\mathbb{R}^d}\frac{\|Ax\|_{2}}{\|x\|_{2}}$}, and $W=\{w_{1},\dots, w_{d}\}$ and $Q=\{q_{1},\dots,q_{d}\}$  are the matrices in  $\mathbb{R}^{n \times d}$ with orthonormal columns.
\par
The next result shows that  with  high probability the subspace which spans the CPAST estimator  \(\widehat{V}(t)\) is close, in terms of \(l,\) to  the subspace spanning $V$ when the number of observation is large enough. We assume that the initial estimator $\widehat{V}(t_0)=\widehat{V}^{0}$ is constructed from $t_0$ first observations by means  of  the singular value decomposition of $\widehat{\Sigma}(t_0)$.

\begin{theorem}
\label{cpast_theorem}
Suppose that the spectral gap condition  \eqref{AD} holds and  
\[
 \sqrt{t_0} \geq    4\sqrt{2} R_{\max}  \frac{\lambda_1+1}{\lambda_d}, 
 \]
where $R_{\max}= 5\sqrt{n-d}+5\sqrt{6}\sqrt{\ln(n\vee T)}$. 
Then after $t-t_0$ iterations   we get
with probability at least $1- {C_0}(n\vee t)^{-2},$   
\begin{equation}
\label{eq:l_bound}
\begin{split}
l (V,\widehat{V}(t)) \leq  & C_1    \frac{\lambda_d+1}{\lambda_d^2} \frac{n-d} {t} +  C_2 \frac{\lambda_1+1}{\lambda_d^2} \frac{\log(n\vee t)}{t},
\end{split} 
\end{equation}
where $C_0,$ \(C_2\) are absolute constants and $C_1$ depends on $\tau.$ 
\end{theorem}
\begin{remark} The second term on the right-hand side of \eqref{eq:l_bound}  corresponds to the error of separating the first $d$  eigenvectors from the rest. The first term is an average error of estimating  all components of $d$ leading eigenvectors. It originates from the interaction of the noise terms with the different coordinates, see \cite{birnbaum2013minimax}.  
\end{remark}
 
\section{Sparse CPAST}
\label{sec: SCPA}
\subsection{Sparsity assumptions on leading eigenvectors}
\label{sparse_case}

We assume that in  the stationary case \eqref{stationary_C} the first $d$ leading eigenvectors $v_i$, $i=1,\dots,d,$ of $\Sigma$ have most of their entries close to zero. Namely, we suppose that each $v_i$ fulfils the so-called weak-$l_r$ ball condition \cite{donoho1993unconditional,johnstone2004sparse}, that is, for some $r\in(0,2),$
\begin{equation*}
|v_i|_{(k)}\leq s_i k^{-1/r},\quad k=1,\dots,n.
\end{equation*}
where $|v_i|_{(k)}$ is the $k$-th largest coordinate of $v_i$. 
The weak-$l_r$ ball condition is known to be more general than $l_r$ ball condition (which is $\|q\|_r\leq s $ for $q\in \mathbb{R}^n$, $r\in (0,2)$, $s\geq 1$), as it combines different definitions of sparsity used in statistics, see \cite{fan2001variable}.
\par
Define a thresholding function $g(x,\beta)$ with a thresholding parameter $\beta>0$ and $x\in\mathbb{R}$ via
\begin{equation}
\label{thresholding_f}
x-\beta \leq g(x,\beta)\leq x+ \beta,\quad g(x,\beta)1_{|x|\leq\beta}=0. 
\end{equation}
For example, the so-called hard-thresholding function $g_H(x,\beta)$  given by 
\begin{equation}
\label{hard}
g_H(x, \beta) = x 1_{(|x|\leq \beta)}
\end{equation}
and the so-called soft-thresholding function  defined as
\begin{equation*}
g_S(x, \beta) =  (x - \beta)_{+}\cdot{\rm sign}(x)
\end{equation*}
fulfill the conditions \eqref{thresholding_f}.
When $\beta$ is a vector with  components $\beta_i$, $i=1,\dots,d$, and $V$ is a matrix with columns $v_i\in \mathbb{R}^n$, $i=1,\dots,d$,  we denote by $g(V,\beta)$  a $n\times d$  matrix with the elements $\{g(v_{ij} ,\beta_i)\}$, $i=1,\dots,d$, $j=1,\dots,n$.

Our primal goal is to propose a subspace tracking method for estimating a $d$-dimensional subspace of the process under  weak-$l_r$ ball   assumption on  the leading eigenvectors  of the covariance matrix $\Sigma$ and  to analyze it's convergence.

\subsection{Initialization and main steps}
Our sparse modification of CPAST relies on the orthogonal iteration scheme with an additional thresholding step   (cf. \cite{ma2013sparse, johnstone2004sparse}). From now on, by a slight abuse of notation, we will denote by $\widehat{V}(t)$ an iterative estimator obtained with the help of the modified CPAST, given $t$ observations.
To get the initial approximation $\widehat{V}(t_0),$ we use the following modification of a standard SPCA scheme, see \cite{ma2013sparse,johnstone2004sparse}.
\begin{enumerate}
\item
First compute the empirical covariance $\widehat{\Sigma}(t_0)$ based on $t_0$ observations:
$
\widehat{\Sigma}(t_0) = \frac{1}{t_0}\sum_{i=1}^{t_0} x(i)x^{\top}(i).
$
\item 
Define a set of indices $G,$ corresponding to large enough diagonal elements of $\widehat{\Sigma}(t_0):$ 
$$G=\left \{k:\,\widehat{\Sigma}_{kk}(t_0)>1+\gamma_0\frac{\log(n\vee t_0 )}{t_0}\right\}$$
for $ \gamma_0 \geq 3\sqrt{2} \frac{\log(n\vee T)}{\log(n \vee t_0)}$. 
\item Let $\widehat{\Sigma}^0(t_0)$ be a submatrix of $\widehat{\Sigma}(t_0)$ corresponding to the row and column indices in  $G\times G$. 
\item As an estimator at  step zero, we take the first $d$ eigenvectors of $\widehat{\Sigma}^0(t_0)$ completed with zeros in the coordinates $\{1,\dots,n\} \backslash G$ to the vectors of length $n$.
\end{enumerate}
Now we describe a sparse modification of CPAST, which we called SCPAST. We start with $\widehat{V}(t_0)$ obtained by the above procedure. 
Then for $t=t_0+1,\dots,T,$  we perform the following steps 
\begin{enumerate}
\item \textit{multiplication}:
$
\widehat{\Upsilon}(t)=\widehat{\Sigma} (t)\widehat{V}(t-1),
$
 \item 
\textit{thresholding}:  define a matrix  $$\widehat{\Upsilon}^{\beta}(t) = g(\widehat{\Upsilon}(t),\beta(t)),$$ where $g$ is a thresholding function satisfying \eqref{thresholding_f} and $\beta(t)$ is the corresponding thresholding vector; 
 \item 
\textit{orthogonalization}:
\begin{equation*}
\widehat{V} (t)=\widehat{\Upsilon}^{\beta}(t) [\widehat{\Upsilon}^{ \beta\top}(t)\widehat{\Upsilon}^{\beta}(t) ]^{-1/2}.
\end{equation*}

\end{enumerate}

%
%
%



\subsection{Convergence of SCPAST }
\label{SCPAST_conv}
First we define the thresholding  parameter  $\beta(t)$ as follows. For $t=t_0+1,\dots,T$ and  $$a  \geq   3\sqrt{2} \frac{\log(n\vee T )}{\log (n\vee t_0)}$$ 
the components $\beta_{i}(t)$, $i=1,\dots,d$ of  the vector $\beta(t)$ are given by 
\begin{equation}
\label{beta_def}
\beta_{i}(t) = a \sqrt{(\lambda_i+1) \frac{\log( n\vee t)}{t}}.
\end{equation}

The motivation for thresholding of the column vectors of $\widehat{\Upsilon}(t)$ comes from the following connection between sparsity of the leading eigenvectors $v_j$, $j=1,\dots,d,$ and the vector $\zeta_v$ with the components $\sqrt{\sum_{j=1}^d \lambda_j v^2_{jk}},$ $ k=1,\ldots, n$ ($\zeta^2_{vk}$ is the variance of the $k$-th coordinate of the signal part   \cite{paul2012augmented}): the weak-$l_r$ sparsity of the vector $\zeta_v$ implies the weak-$l_r$ sparsity of $v_j$, $j=1,\dots,d$. 
\par 
Suppose that  $d$  and the eigenvalues $\lambda_1,\dots,\lambda_{d}$ are known. 
 In the  case of unknown  $d$ and  $\lambda_1,\dots,\lambda_{d}$  one might first estimate the eigenvalues of $\widehat{\Sigma}^{0}(t_0)$ defined in the previous section and then select the largest set of eigenvalues satisfying the spectral gap condition \eqref{AD}  with some parameter ${\tau}$   (see~\cite{ma2013sparse} for more details). 
\par
Denote by $S(t)$ the set of indices of ``large'' eigenvectors components ($S$ stands for "signal"), that is,  for  a fixed $t,$   
$$
S(t) =\left\{j:  |v_{ij}|\geq b h_i \sqrt{ \frac{\log(n\vee t)}{t}},\,\text{for some } i=1,\dots,d\right\},
$$
where  $h_i =  \frac{\sqrt{\lambda_i +1}}{\lambda_i}$ and
$ b = \frac{0.1 a}{\sqrt{\tau} \sqrt{d}}.$  In fact, the quantity $h^2_i/t$ is an estimate   of the noise variance in the  entries of the $i$-th leading eigenvector \cite{birnbaum2013minimax}. The number of ``large'' entries of the first $d$ leading eigenvectors  to estimate thus might be estimated by the cardinality of $S(t)$, which we denote by ${\rm card}(S(t))$. 
One can bound  ${\rm card}(S(t))$ 
as 
$$
{\rm card}(S(t)) \leq \sum_{i=1}^d {\rm card}(S_j(t)),  
$$
where $S_j(t)) = \left\{j:  |v_{ij}|\geq b h_i \sqrt{ \frac{\log(n\vee t)}{t}} \right\}$. 
From Lemma \ref{eff_dim} (see Appendix B) we see that
 $
d \leq  {\rm card}(S(t))\leq C M(t),
 $
 where $C$ depends on $b,$  $r$ and  
 \begin{equation}
 \label{M}
 M(t)= n\wedge \left[\sum_{j=1}^d \frac {s_j^r}{h_j^{r} }\left( \frac{\log(n\vee t)}{t}\right)^{-r/2} \right].
 \end{equation} 
Note that  in the sparse case, the number of non-zero components ${\rm card}(S(t))$ is much smaller than $n$.
For  example, if  $\|v_j\|_r\leq s$, $j=1\dots,d,$ then 
$$M(t) \leq  n \wedge  d\frac{s^r}{h_d^{r}} \left(\frac{\log(n\vee t)}{t}\right)^{-r/2}.$$  The value $\frac{s^r}{h_j^{r}}$ is often referred to as an effective dimension of the vector $v_j$. Thus $M(t)$ is the number of effective coordinates of $v_j$, $j=1,\dots,d$ in the case of disjoint $S_j(t)$.

Since $ {h_d}^2/t$ is an upper-bound for the estimation error for the components of the first $d$ leading eigenvectors, the right hand side of the above inequality gives,  up to a logarithmic term, the overall number of components of the $d$ leading eigenvectors to estimate. The next theorem gives non-asymptotic bounds for a distance between \(V\) and
\(\widehat{V} (t).\)
 

\begin{theorem}
\label{scpast_theorem}

Let 
\begin{equation}
\label{t0_sparse}
\sqrt{t_0}\geq    
 \left(C_1  h_d M^{1/2}(T) + C_2 
  \right)\frac{  \lambda_{1}+1 }{ \lambda_d}\sqrt{ \log(n\vee T)}, 
\end{equation} 
where $C_1$ depends on $\tau$ in \eqref{AD}, $r$, $a$,  $C_2$ depends on $\tau$. %
After $t$ iterations one has with probability at least $1-C_0(n\vee   t)^{-2},$ 
\begin{equation}
\label{eq:l_bound_s}
\begin{split}
l (V,\widehat{V} (t)) \leq  & C_1    h_d^2 M(t)    \frac{\log(n\vee t)}{t}  +  C_2 \frac{\lambda_1+1}{ \lambda_d ^2} \frac{\log(n\vee t)}{t}.
\end{split} 
\end{equation}
with some absolute constant \(C_0>0.\)
\end{theorem}
 
%
 
\begin{remark}  The second term in \eqref{eq:l_bound_s}  is the same as in the non-sparse case, see Theorem \ref{cpast_theorem}. This term  is always present as an error of  separating the first $d$ eigenvectors from the rest eigenvectors regardless  how sparse they are. The first term in \eqref{eq:l_bound_s} and \eqref{eq:l_bound} is responsible for the interaction of the noise with different coordinates of the signal. The average error of estimating  one entry of the first $d$ leading eigenvectors based on $t$ observation can be  bounded by $\frac{1}{t} \frac{\lambda_d+1}{\lambda_d^2} ,$ see \cite{paul2004nonparametric}. The number of components to be estimated in SCPAST for each vector is bounded by $M(t)$  (see \eqref{M}), which is small  compared to \(n\) in the sparse case. Thus, the first term in \eqref{eq:l_bound_s} can be significantly smaller than the first one in \eqref{eq:l_bound}, provided the first $d$ leading eigenvectors are sparse. Note also that the computational complexity of SCPAST at each step $t=t_0+1,\dots,T$ is $O(d^2 {\rm card} (S(t)))$  with probability given by Theorem \eqref{scpast_theorem}. 
\end{remark}

\section{Numerical results} 
\label{sec:sim}
\subsection{Single spike}
\begin{figure}
\begin{multicols}{2}
\begin{center} {\tiny
(a)

\includegraphics[width=0.44\textwidth, height = 0.24\textheight]{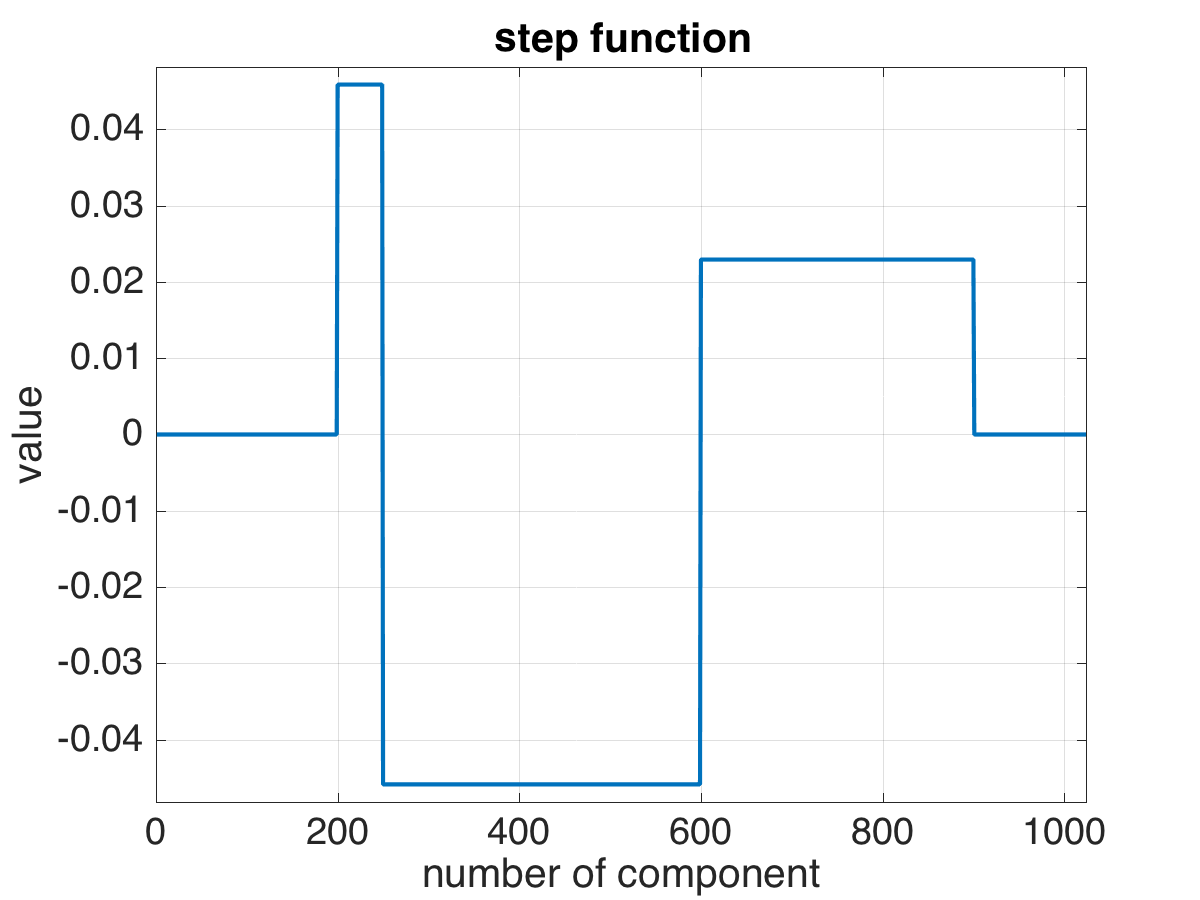}

(b)

\includegraphics[width=0.44\textwidth]{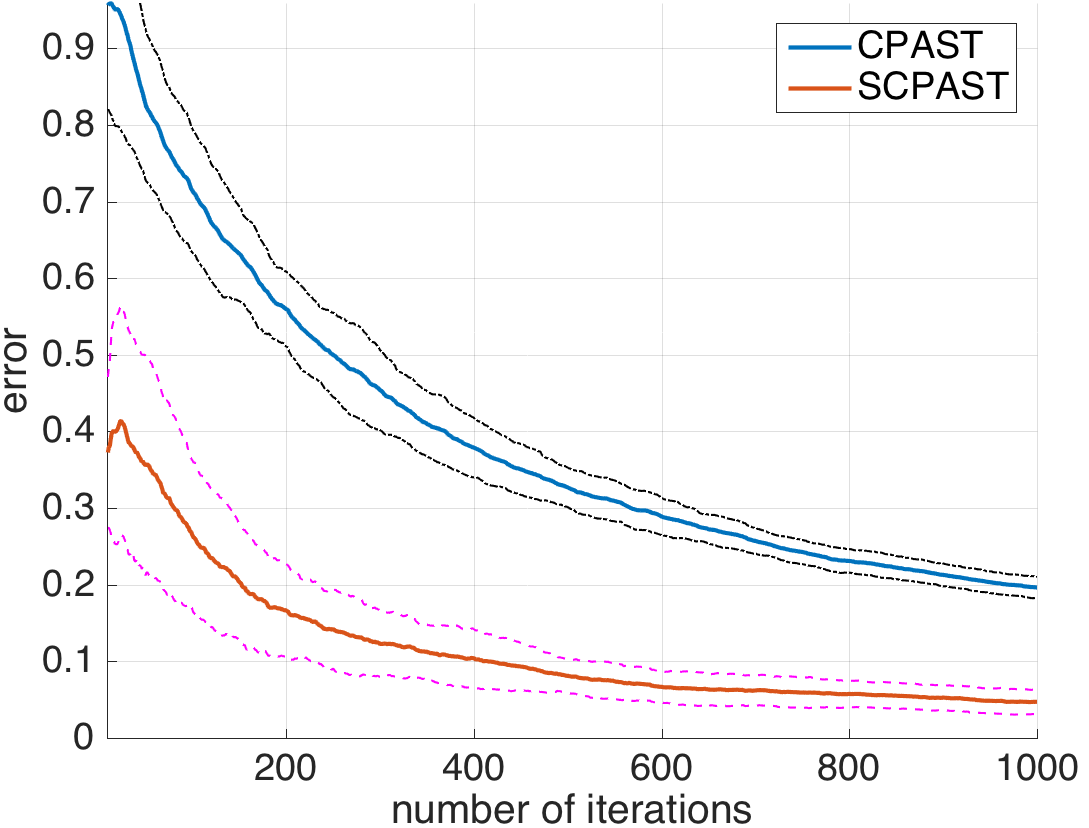}

(c)

\includegraphics[width=0.44\textwidth]{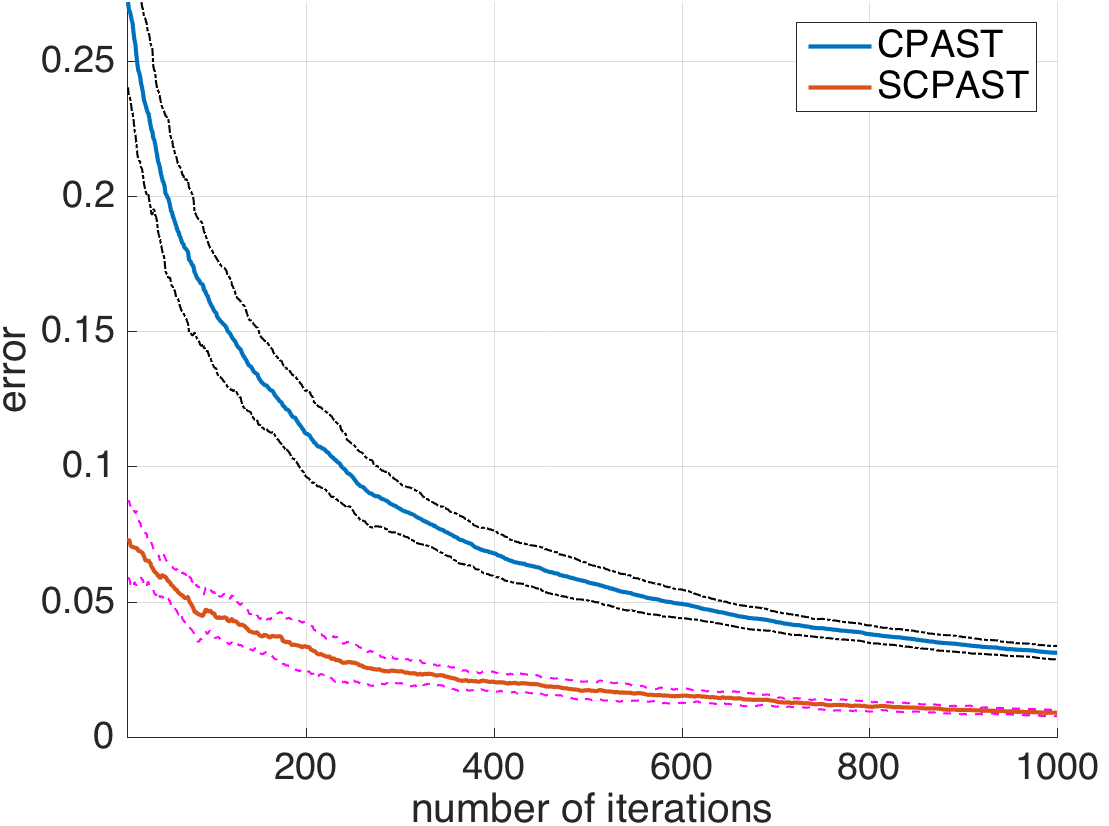}

(d)

\includegraphics[width=0.44\textwidth]{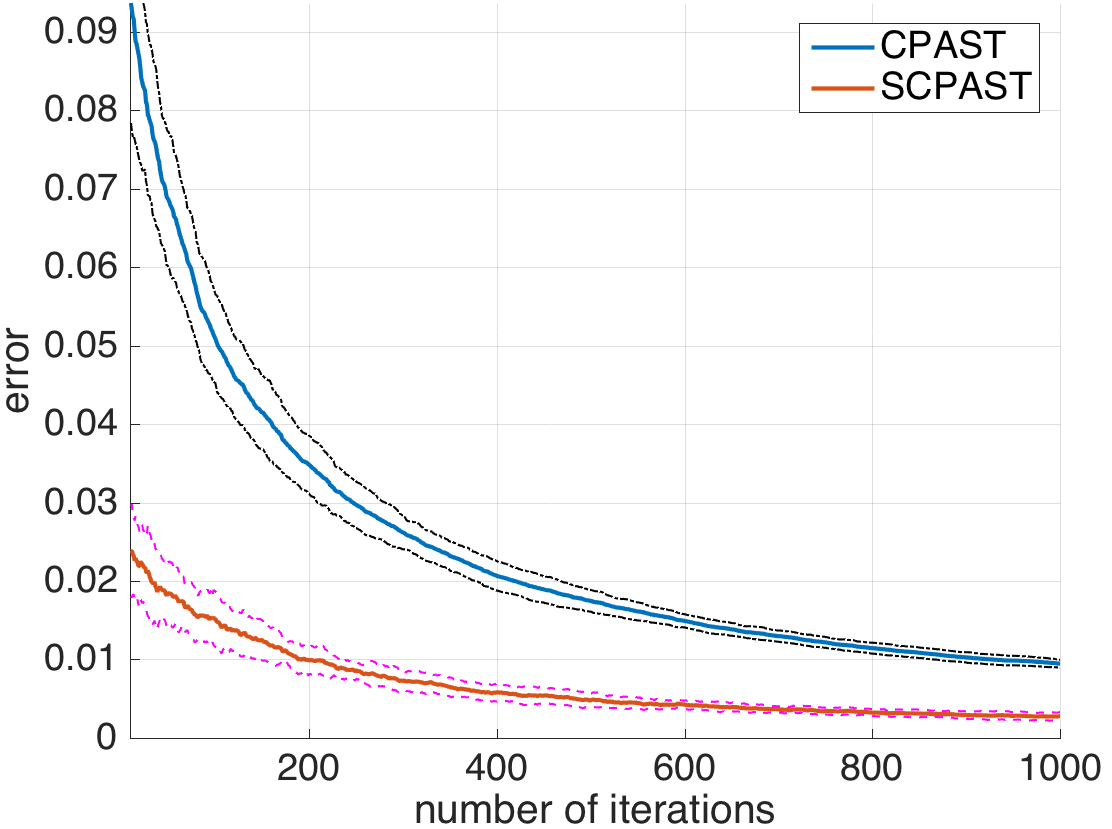}
}
\end{center}
\end{multicols} 
\caption{\label{pic1} The components of the leading eigenvector to recover (a) step function, (b)--(d) contain the results for the error $l(v_1,\widehat{v}_1)$ for  $\lambda_1 = \{5,30,100\}$} 
\end{figure}

\begin{figure}
\begin{multicols}{2}
\begin{center} {\tiny
(a)

\includegraphics[width=0.44\textwidth, height = 0.24\textheight]{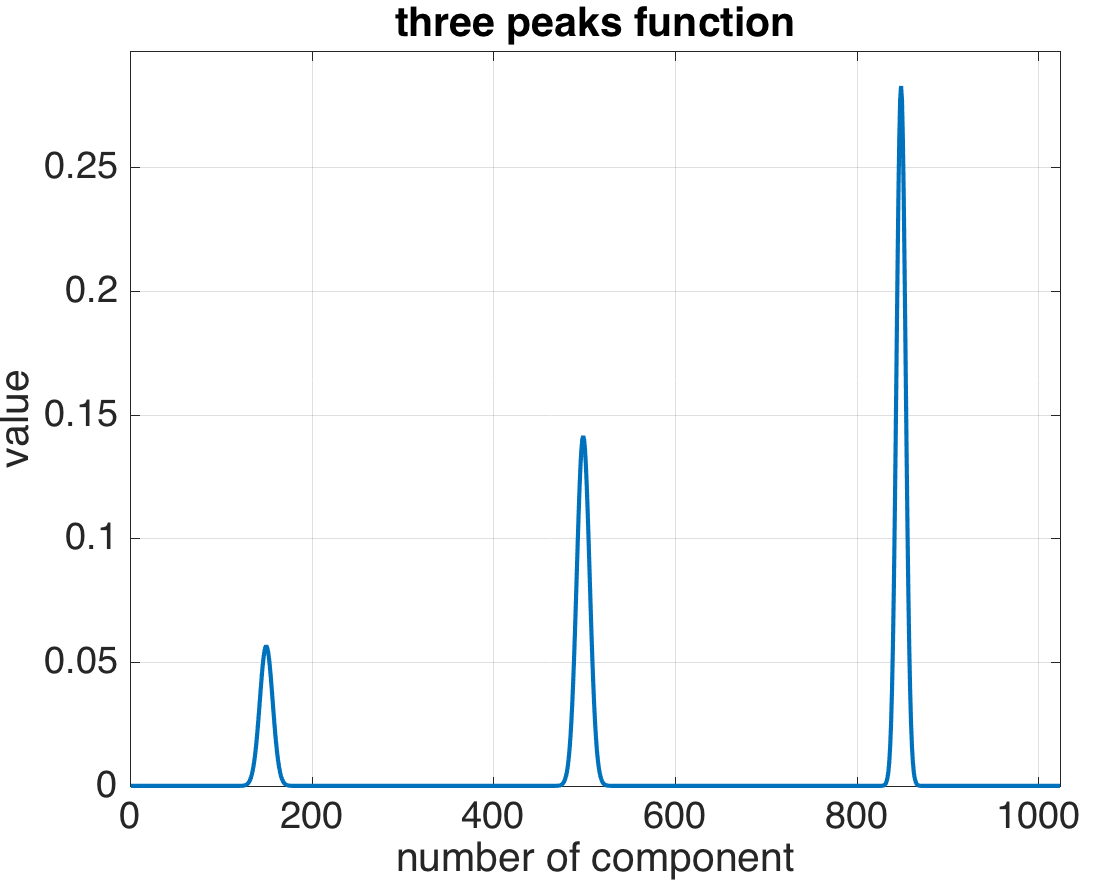}

(b)

\includegraphics[width=0.44\textwidth]{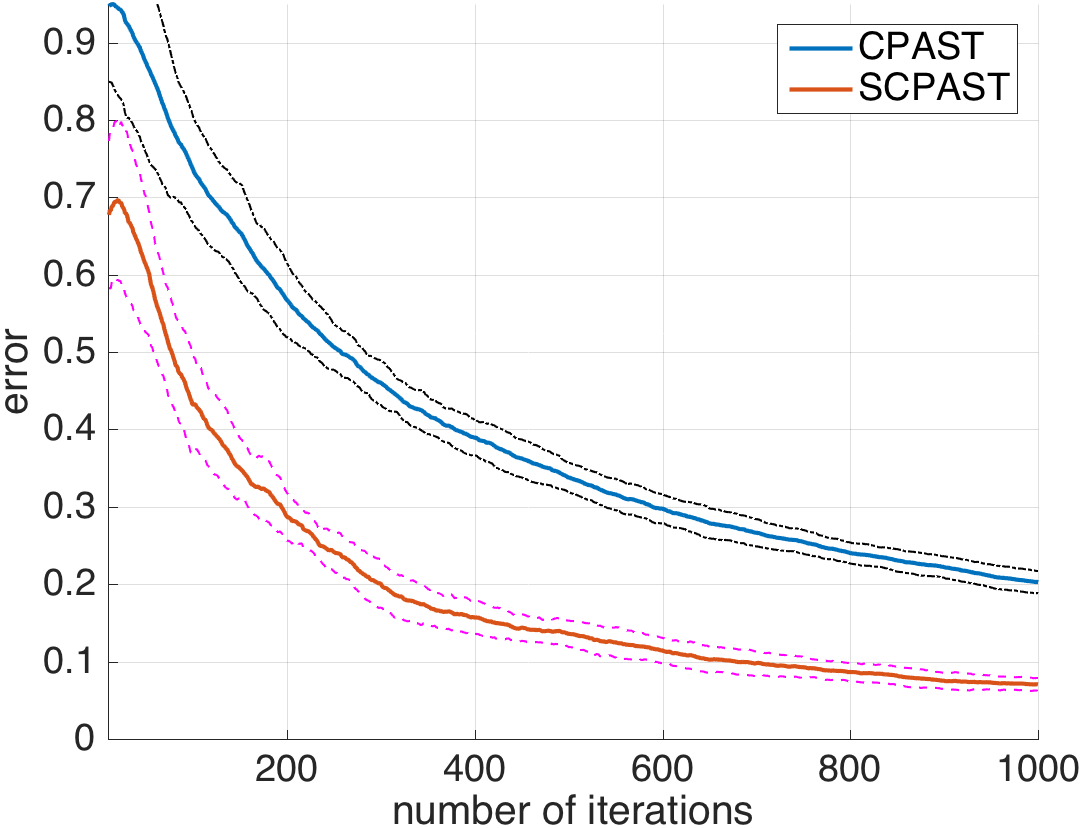}

(c)

\includegraphics[width=0.44\textwidth]{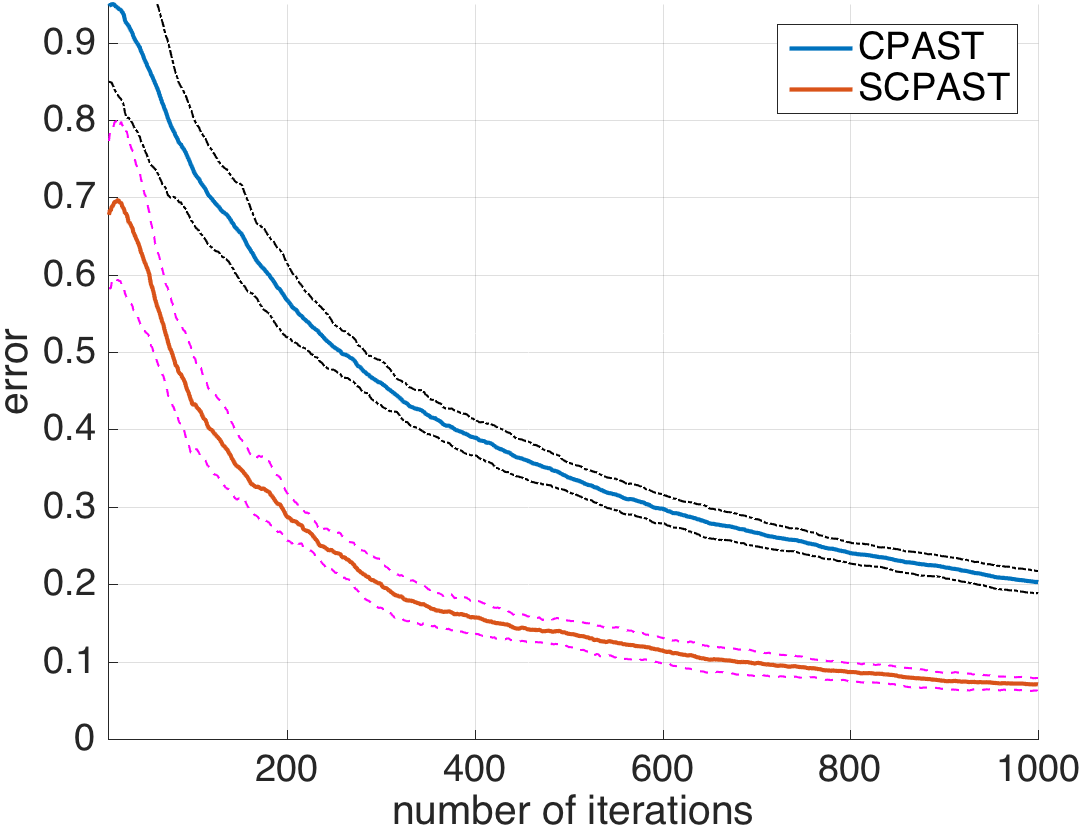}

(d)

\includegraphics[width=0.44\textwidth]{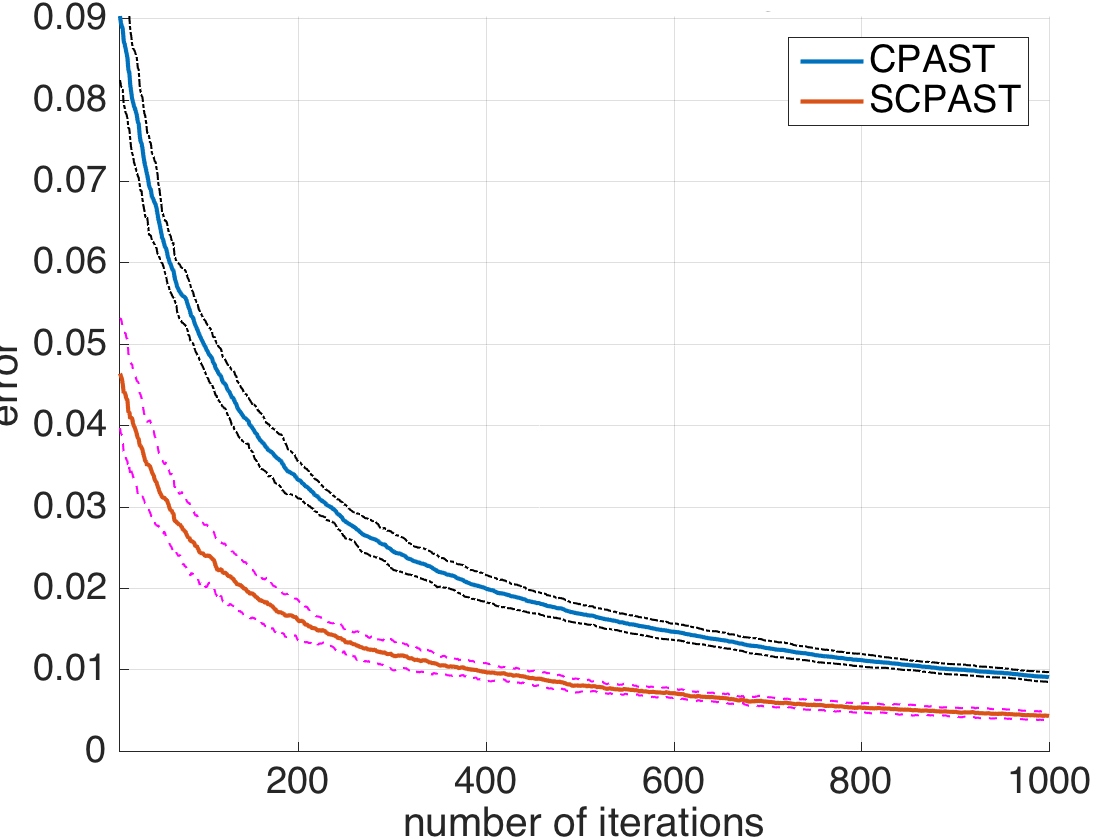}
}
\end{center}
\end{multicols}
\caption{\label{pic2} The components of the leading eigenvector to recover (a) three peeks function, (b)--(d) contain the results for the error $l(v_1,\widehat{v}_1)$ for  $\lambda_1 = \{5,30,100\}$} 
\end{figure}

\begin{figure}
 \begin{multicols}{2}[\columnsep=0.01cm]
 {\tiny
 \begin{center}
(a)

\includegraphics[width=0.44\textwidth, height = 0.24\textheight]{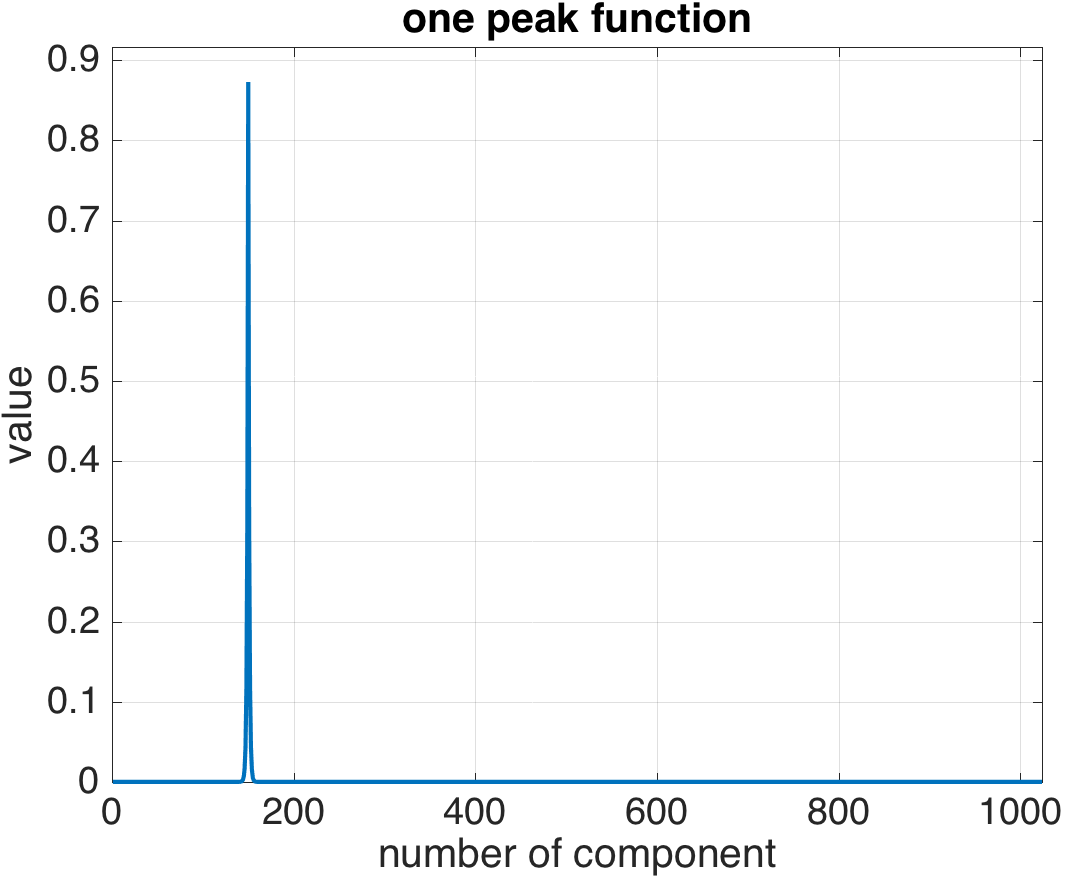}

(b)

\includegraphics[width=0.44\textwidth]{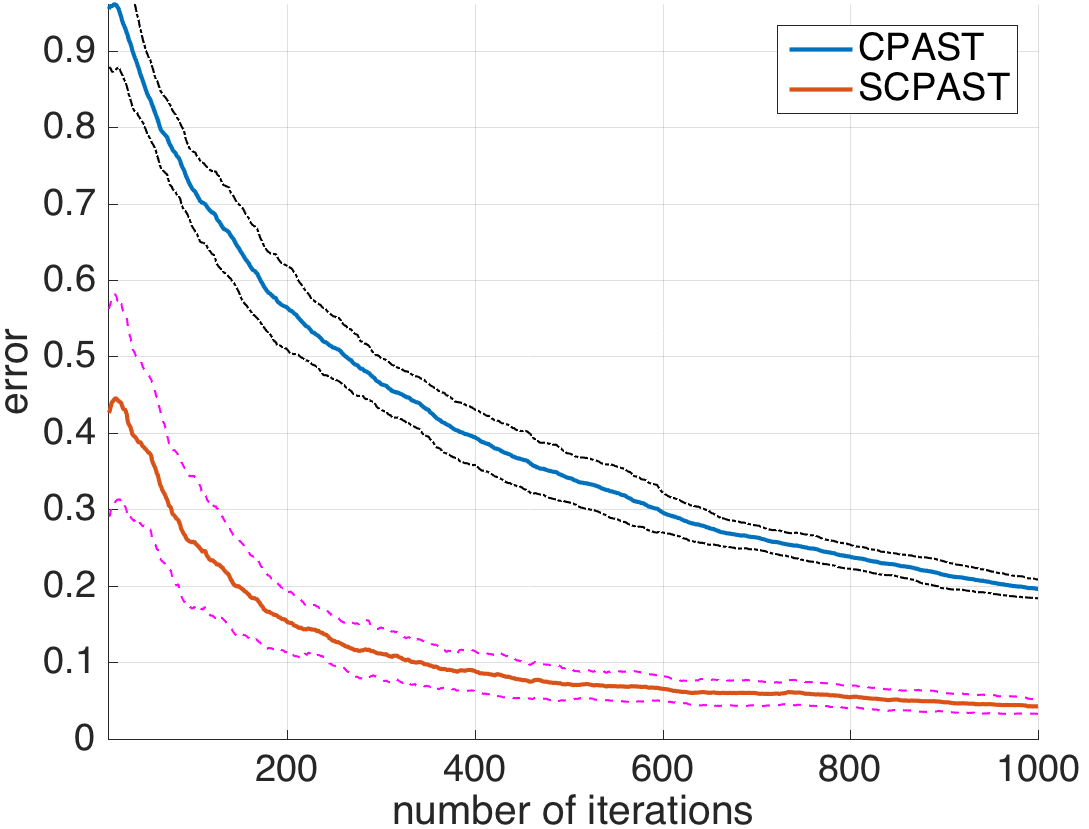}

(c)

\includegraphics[width=0.44\textwidth]{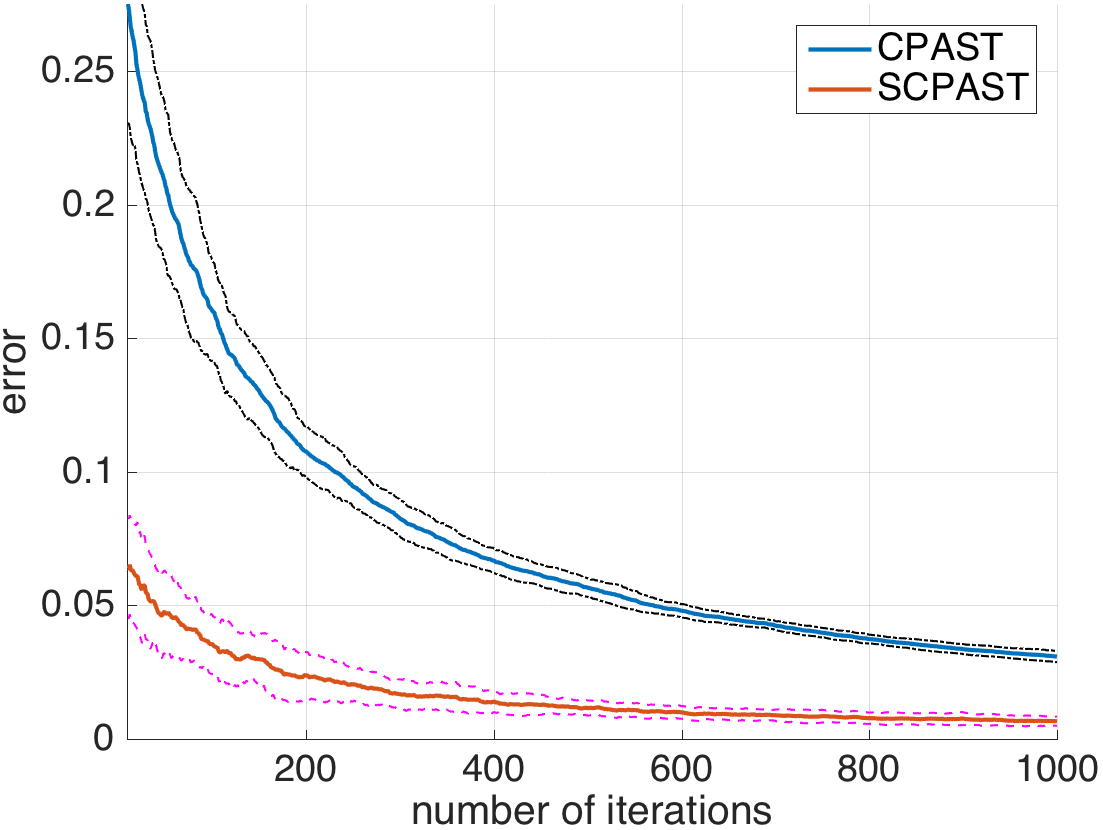}

(d)

\includegraphics[width=0.44\textwidth]{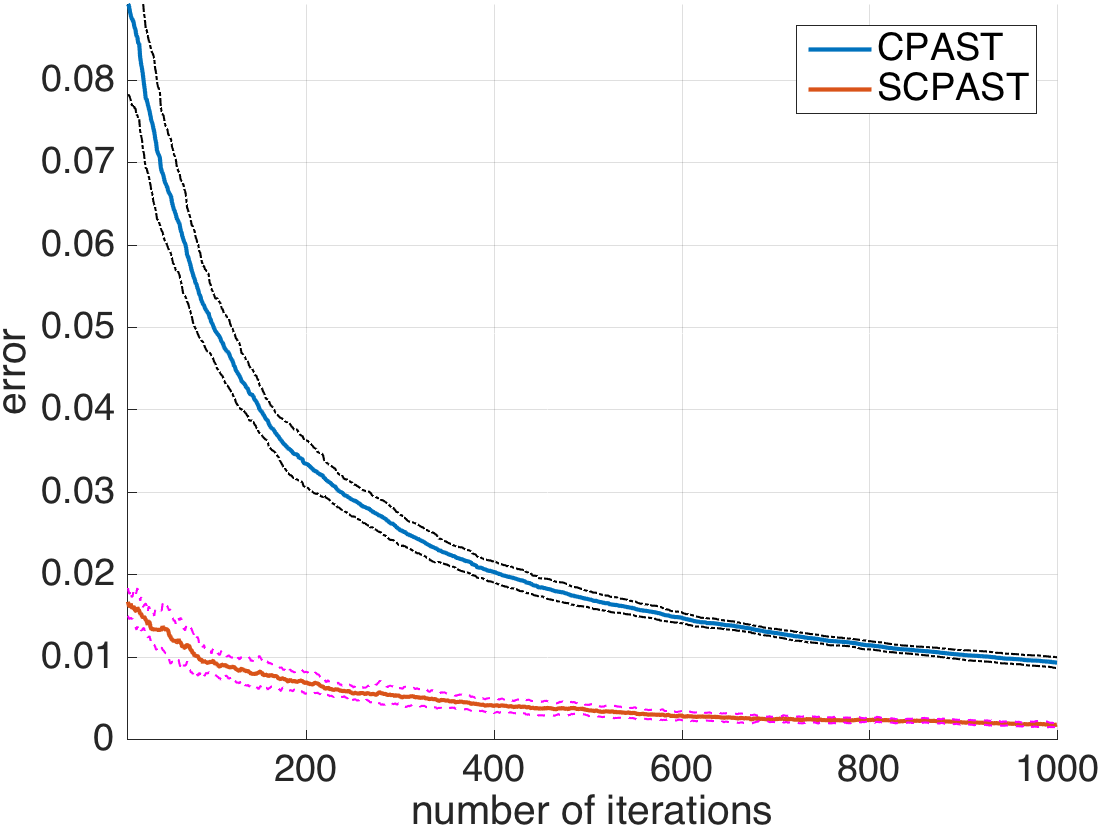}
\end{center}
} 
\end{multicols}
\caption{\label{pic3} The components of the leading eigenvector to recover (a) one peek function, (b)--(d) contain the results for the error $l(v_1,\widehat{v}_1)$ for  $\lambda_1 = \{5,30,100\}$} 
\end{figure}

 To illustrate the advantage of using SCPAST  for the sparse case, we generate $T=2000$ observations from  \eqref{stationary_model} for the case of a single spike, that is, $d=1$ and $n=1024 $. Our aim is to estimate the leading eigenvector $v_1$. 
We shall use  three functions depicted in subplots (a) of Fig. \ref{pic1}--\ref{pic3}  with  different sparsity levels in the wavelet domain.

 The observations are generated for the noise variance $1$ and following cases of maximal eigenvalue  $\lambda_{1} \in \{5, 30,100\}$. We used the Symmlet 8 basis from the Matlab package SPCALab to transform the initial data into the wavelet domain.
We applied CPAST and SCPAST for the recovery of wavelet coefficients of the vector $v_1$ and then transformed the estimates to the initial domain and computed the error $l(v_1,\widehat{v}_1)$ depending on the number of observations. The results for the hard thresholding \eqref{hard} with the $a=1.5$ are shown in Fig. \ref{pic1}-\ref{pic3} in subplots (b)-(d).    Note that one peak function has sparser wavelet coefficients than those of three peak functions and the error of the recovery with SCPAST is significantly smaller for the  case of one peak function.
 
\subsection{Real data example}
\begin{figure}[h] 
\centering
\includegraphics[angle=0, width=0.7\linewidth]{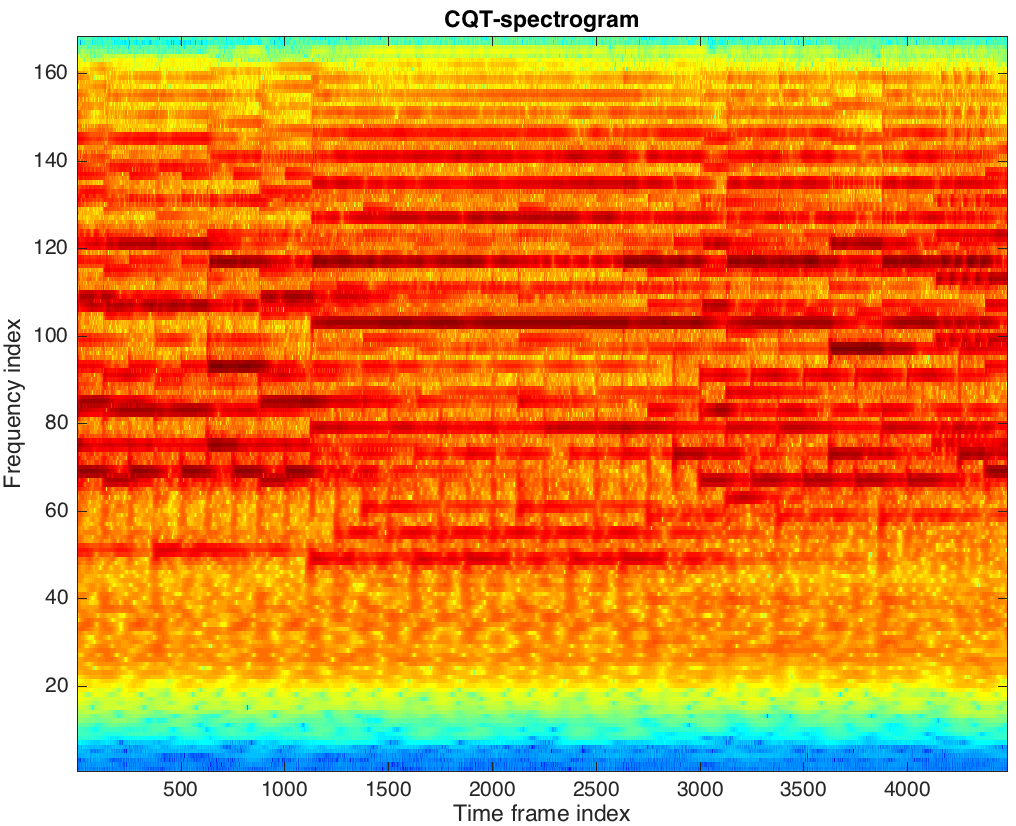}
\caption{\label{fig_sp1} CQT-Spectrogram of Bach Siciliano for Oboe and Piano. The deep blue color corresponds to the zero values, the red color corresponds to the higher values.} 
\end{figure}

Natural acoustic signals like the musical ones exhibit a highly varying temporal structure, therefore there is a need in adaptive unsupervised methods for signal processing which reduce the complexity of the signal.     
In  \cite{krymova2017segmentation} a method was proposed which reduces the spectral complexity of music signals
using the adaptive segmentation of the signal in the spectral domain for the principal component analysis for listeners with cochlear hearing loss.   
In the following we apply  CPAST and SCPAST as an alternative method for the complexity reduction of music signals.
To illustrate the use of SCPAST and CPAST we set the memory parameter $\gamma=0.9$  to be able to adapt to the changes in the spectral domain of the  signal.  We focus on the first leading eigenvector recovery. As an example we consider  a piece from Bach Siciliano for Oboe and Piano.  
A  wavelet-kind CQT-transform \cite{brown1991calculation} 
 is computed for the signal (see a spectrogram of the transform in Fig. \ref{fig_sp1}).   
The warmer colors correspond to  the higher  values of the amplitudes of the harmonics present in the signal at a particular time frame.  It is clear that the signal has some regions of  ``stationarity'' (e.g. approximately in time frame interval $[1200,2600]$). 
We regard the corresponding spectrogram as a matrix with $4500$ observations of $168$-dimensional signal modeled by \eqref{data_repres}  and apply SCPAST and CPAST methods  to recover the leading eigenvector $v_1$. 

\begin{figure}[h]
\begin{multicols}{2}
\includegraphics[width=0.5\textwidth]{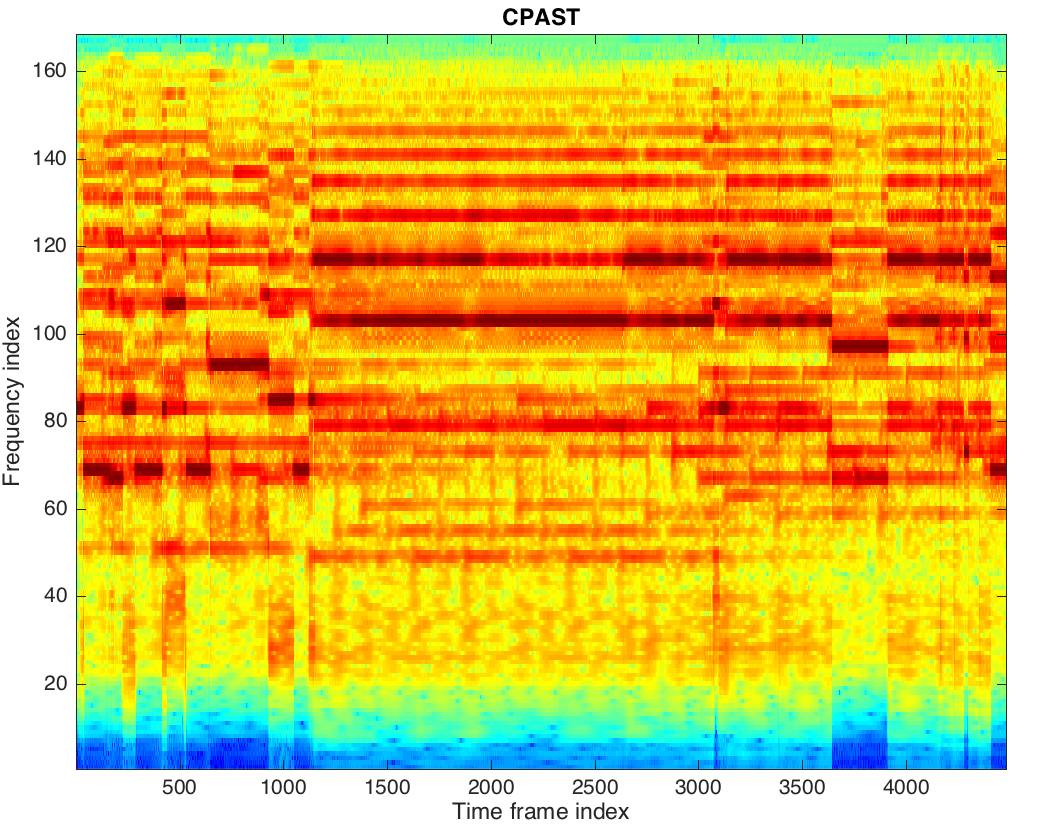}

\includegraphics[width=0.5\textwidth]{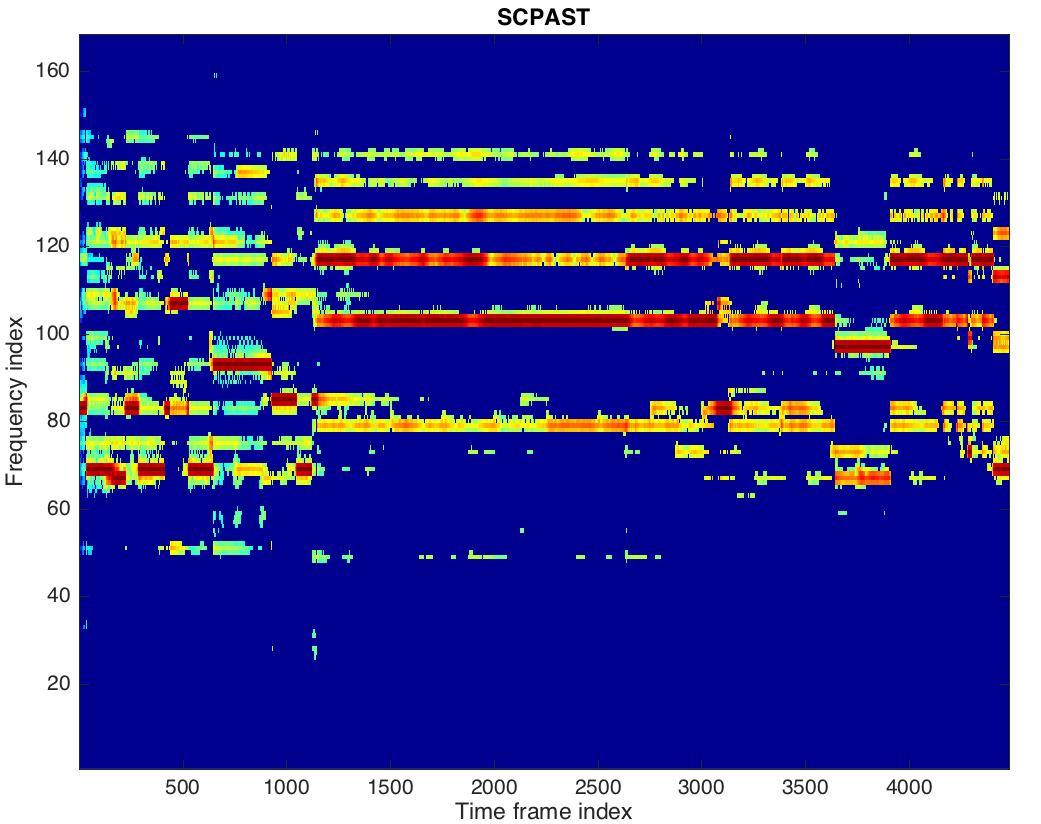} 
\end{multicols}
\caption{\label{fig_sp2} CQT-Spectrogram of the leading eigenvector recovered by  CPAST and SCPAST  with the memory parameter $\gamma=0.9$ for Bach Siciliano for Oboe and Piano.} 
 \end{figure}
  Fig. \ref{fig_sp2} contains the results of the  recovery  of the leading eigenvalue with $168$ components.  
The results show that SCPAST method allows to obtain sparse representation of the leading eigenvectors and seems to be promising for construction of the structure-preserving compressed representations of the signals. 

\section{Sketch of the proofs}
\label{sec:proofs}
Denote   by $\bar{V}$ a matrix with $n-d$
column  vectors     $v_i$, $i=d+1,\dots, n,$ which complete the orthonormal columns $\{v_i\}_{i=1}^d$ of the matrix $V$ to the orthonormal basis in $\mathbb{R}^{n}$.
Denote by $X(t)$ a matrix with the columns $\{x(i)\}_{i=1}^t$. From \eqref{stationary_model} one gets a representation 
\begin{equation}
\label{data_repres}
X(t) =V \Lambda_d^{1/2}  U^{\top}(t) + \sigma \Xi(t), \quad t =1,\dots,T,
\end{equation}
where $U(t)\in \mathbb{R}^{t\times d}$, $\Xi(t) \in \mathbb{R}^{n\times t}$ are  matrices  with  independent $\mathcal{N}(0,1)$ entries, $V $ is the orthonormal matrix with columns $\{v_i\}_{i=1}^n$, $\Lambda_d$  is a diagonal matrix  with  $\lambda_i$, $i=1,\dots,d$ on the diagonal. Denote a set of indices  to  the small components of leading eigenvectors as $N(t)=\{1,\dots,n\}\backslash S(t)$  (where $N$ here stands for ``noise''). 

From \eqref{data_repres} the empirical covariance matrix can be decomposed as 
\begin{equation}
\label{cov_decomp}
\begin{split}
\widehat{\Sigma}(t) 
= & \frac{1}{t} V \Lambda^{1/2}_d U^{\top} (t)U(t)\Lambda^{1/2} _dV^{\top} + \frac{1}{t} \Xi(t)\Xi^{\top}(t) \\
 &+  \frac{1}{t} V \Lambda^{1/2}_d U^{\top} (t)\Xi^{\top} (t) + \frac{1}{t}\Xi(t)U(t)\Lambda^{1/2}_dV^{\top}.
\end{split}
\end{equation}
It is well known \cite{golub2012matrix}  that the distance \eqref{distance} between  subspaces $\mathcal{W}$ and $\mathcal{Q}$, spanning $n\times d$ matrices with orthonormal columns $W$ and $Q$ correspondingly,
is related  to $d$-th principal angle between subspaces $\mathcal{W}$ and $\mathcal{Q}$
as $ 
l(\mathcal{W},\mathcal{Q})= \sin^{2}\phi_{d}( \mathcal{W}, \mathcal{Q}),
$
where the principal angles  $0\leq\phi_{1}\leq\dots\leq\phi_{d}$ 
between subspaces $\mathcal{W}$ and $\mathcal{Q}$ are recursively defined as  \cite{hardt2014noisy}
\begin{equation*} 
\begin{split}
\phi_{i}(\mathcal{W},\mathcal{Q})&= \arccos\frac{\langle x_i,y_i\rangle}{\|x_i\|_{2}\|y_i\|_{2}},\,\text{where}
\\
\{x_i,y_i\}&=\argmin\limits_{\substack{x\in\mathcal{W},\,y\in\mathcal{Q},\\ x\perp x_{j},\,y\perp y_{j},\,j<i}}\biggl\{\arccos\frac{\langle x,y\rangle}{\|x\|_{2}\|y\|_{2}}\biggr\}.
\end{split}
\end{equation*}
From the variational characterization of the singular values and the above definition of the principal angles,    the $d$-th principal angle between subspaces spanning the columns of  $W$ and $Q$ has the following non-recursive definition 
\begin{align} 
\label{cos}
\cos\phi_{d}( {W},  {Q})=& 
 = \min_{\|x\|_{2}=1,\,x\in\mathbb{R}^d}\frac{\|W^{\top}Qx\|}{\|Qx\|},\\
\label{tan}
\tan\phi_{d}( {W}, {Q})=&  \max_{\|x\|_{2}=1,\,x\in\mathbb{R}^d}\frac{\|\bar{W}^{\top}Qx\|}{\|W^{\top}Qx\|}.
\end{align}
In the next sections we derive the error bounds for CPAST and SCPAST  by looking at the change of the $d$-th principal angle between the  eigensubspace spanning the columns of $V$ and its estimators based on $t$ observation, where $t=t_0+1,\dots,T$.
\subsection{Bound for CPAST } 
\label{cpast}
The aim of this section is to show that  with the high probability the subspace which spans CPAST estimator $\widehat{V}(t)$ \eqref{cpast} is close to  the subspace, which spans $V$ when the number of observations is large enough. We assume that the initial estimator $\widehat{V}^{0}$ is constructed from first $t_0$  observations with the help of SVD of $\widehat{\Sigma}(t_0)$. Let us first state   the bound for the error  $l (\widehat{V}(t),V)$ which depends on the error on the previous iteration $l (\widehat{V}(t-1),V)$ for the fixed $t=t_0+1,\dots,T$.
Denote $r(t)= l^{1/2}(\widehat{V}(t),V) = \sin\phi_{d}(\widehat{V}(u),V)$.
\begin{lemma}\label{lemma_1}
For CPAST \eqref{cpast_1} with \mbox{probability $1-C_0 (n\vee t)^{-3}$}
\begin{equation}
\label{tan_iter}
\begin{split}
r(t)  \leq&\frac{ (\lambda_{d+1}+1)\tan\phi_{d}(\widehat{V}(t-1),V)}{\lambda_{d}+1-(\lambda_{1}+1 )E (t ) \sec\phi_{d}(\widehat{V}(t-1),V)   } \\
 & +\frac{ (\lambda_{1}+1 )^{1/2} E (t ) \sec \phi_{d}(\widehat{V}(t-1),V)}{\lambda_{d}+1-(\lambda_{1}+1) E (t) \sec \phi_{d}(\widehat{V}(t-1),V)},
\end{split}
\end{equation}
where 
$$E( t) =  5 \sqrt{\frac{n-d}{t}} + 5\sqrt{6}  \sqrt{\frac{\log(n\vee t)}{t}}.$$
\end{lemma}
The following lemma gives the bound for the   error  $l (\widehat{V}(u),V)$ depending on the error of the previous iteration $l (\widehat{V}(u-1),V)$ for all $u\in \{t_0+1,\dots,t\}$. 
\begin{lemma} 
 \label{lemma_2}
 With probability greater than 
 $$1-C_0(n\vee t)^{-3}$$ 
  for all $u\in\{t_0+1,\dots,t\}$
\begin{equation}
\label{iter_alg_0}
r(u) \leq \frac{\alpha_{0}r(u-1)+\alpha_{1}\frac{R(t)}{\sqrt{u}} }{\sqrt{1-r^{2}(u-1)}-\alpha_{2}\frac{R(t)}{\sqrt{u}} }, 
\end{equation} 
where $r(u) =  \sin\phi_{d}(\widehat{V}(u),V)$,  
$$
R(t) = 5 \sqrt{n-d} +5\sqrt{6} \sqrt{\log(n\vee t)}, 
$$  
\begin{equation}
\begin{split}
\label{alphas0}
\alpha_0 = \frac{1 }{\lambda_d+1},\quad \alpha_1 = \frac{\sqrt{\lambda_1+1}}{ {\lambda_d+1}},\quad
 \alpha_2 = \frac{\lambda_1+1}{\lambda_d+1}.
\end{split}
\end{equation}
\end{lemma} 
Given Lemma \ref{lemma_2} it is possible to derive a  bound for $l (\widehat{V}(t),V)$. First let us state the result which allows to bound the error of the initial estimate $\widehat{V}(t_0)$.
\begin{lemma}
\label{init}
Let $\widehat{V}(t_0)$ be a matrix containing first $d$ leading eigenvectors of the matrix $\widehat{\Sigma}(t_0)$. Then with probability $1-C_0 (n\vee T)^{-2}$
$$
r^2(t_0)=l(\widehat{V}(t_0),V)\leq \alpha^2 \frac{1}{t_0},
$$
where
$\alpha=  R_{\max} \frac{\lambda_1+1}{\lambda_{d}},$
 with $R_{\max}=R(T)$.
\end{lemma}
The following Lemma gives the error of CPAST after observing $K$ vectors $x_i$, $i=t_0+1,\dots,t_0+K$  based on the recursive bound \eqref{iter_alg_0}. Note that the proof of Lemma \ref{lemma_3} also insures that the denominator in \eqref{iter_alg_0} is bounded  away from zero.   
\begin{lemma}
\label{lemma_3}
Suppose that 
$r(t_0)\leq  \alpha \frac{1}{\sqrt{t_0}}$, 
where 
\begin{equation*}
 \sqrt{t_0} \geq    8 R_{\max} \frac{\alpha_{2}}{ (1-\alpha_0)^{3/2}} .
\end{equation*}
Then for $K\geq K(\rho,t_0)$
\begin{equation}
\label{r_0_bound}
r(t_{0}+K)\leq 2 \frac{\alpha_{1}}{\alpha_{0}}\frac{1}{1-\alpha_0} \frac{R(t_0+K)}{\sqrt{K+t_{0}}}.
\end{equation}
\end{lemma}
The statement of the Theorem \ref{cpast_theorem} follows from Lemma \ref{lemma_2} applied to the inequality  \eqref{iter_alg_0}  with \eqref{alphas0}, which holds with probability 
$1- C_{0} (n\vee t)^{-2}$ with the initial conditions given by Lemma  \ref{init}. 
 
\subsection{Bound for SCPAST}
Define   $\widehat{\Sigma}^{\circ}(t)$, the oracle version  of $\widehat{\Sigma}(t)$ and the corresponding expectation ${\Sigma}^{\circ}(t)  = \mathbf{E} \widehat{\Sigma}^{\circ}(t)$ as follows 
 \begin{equation}
\label{mat_break}
\widehat{\Sigma}^{\circ}(t)=
 \left[\begin{matrix}
 \widehat{\Sigma}_{S}(t) & 0\\
 0 & I_{N(t)}
  \end{matrix}\right]
,\,
 {\Sigma}^{\circ}(t)  = 
\left[\begin{matrix}
 \Sigma_{S}(t) &0\\
0 & I_{N(t)}
 \end{matrix}
 \right],
 \end{equation}
 where  $\widehat{\Sigma}_{S(t)}$ and $\Sigma_{S}(t)$ are  the sub-matrices of the size ${\rm card}(S(t))\times {\rm card} (S(t))$ with column and row indices from $S(t)$. The identity matrix $I_{N(t)}$ has the size $ {\rm card}{(N(t))}\times  {\rm card}{(N(t))}$. 
 Here we assumed without loss of generality that indices in $ {S(t)}$ are always smaller than ones in $N(t)$.  

First we  obtain the oracle sequence $\widehat{V}^{\circ}(t)$ of the solutions by iterating SCPAST with matrices $\widehat{\Sigma}^{\circ}(t)$ instead of $\widehat{\Sigma} (t)$. We define the initial estimate $\widehat{V}^{\circ}(t_0)$  with the steps (a)-(d) in the section \ref{cpast} applied to the matrix $\widehat{\Sigma}^{\circ}(t_0)$. And then bound $\sin\phi_d (\widehat{V}^{\circ}(t),V  )$.  
 Denote the result of the thresholding  the columns of the matrix $\widehat{\Upsilon}^{\circ}(t) =\widehat{\Sigma}^{\circ}(t)\widehat{V}(t-1)$ with the thresholding parameters given by the vector $\beta(t)$   as  $$\widehat{\Upsilon}^{\circ, \beta}(t) = g(\widehat{\Upsilon}^{\circ}(t), \beta(t)).$$  

Denote the submatrix $V_{S}(t)$ obtained by selecting the rows of $V$ with indices in $S(t)$. Denote  by $V_k$, $k=1,\dots,n$   the rows of $V$ (recall that the columns are $v_j$, $j=1,\dots,m$). For the estimators of $V$ we omit the dependence of $S(t)$ on $t$ as the estimator itself depends on $t$, that is, $\widehat{V}_S(t)$ is a  matrix of the rows of $\widehat{V}(t)$ with indices from $S(t)$.

The following  bound for the oracle error $r^2(t)  =l(\widehat{V}^{\circ}(t),V )$ of   SCPAST method is analogous to Lemma \ref{lemma_2}. 
\begin{lemma}
\label{lemma_5}
For $u=t_0+1,\dots,t$   
 with probability greater than 
 $1-C_0(n\vee t)^{-3}$ 
 the following  bound  holds true  
\begin{equation}
\label{iter_alg_0s}
r(u) \leq \frac{\alpha_{0}r(u-1)+\alpha_{1}\frac{R^{\circ}(t)}{\sqrt{u}} }{\sqrt{1-r^{2}(u-1)}-\alpha_{2}\frac{R^{\circ}(t)}{\sqrt{u}} },\quad \text{where }
\end{equation} 
\begin{equation}
\begin{split}
\label{alphas}
\alpha_0 &= (\lambda _d +1)^{-1},\quad \alpha_1 =\alpha_2 = (\lambda_1+1)\slash (\lambda_d+1), 
\end{split}
\end{equation}
$
R^{\circ}(t) =C_1 h_d M^{1/2}(t)\sqrt{ \log(n\vee t)} + C_2\sqrt{  \log(n\vee t)}
$, where  $C_0$ and $C_2$ are constants and $C_1$ depends on $r$, $d$, $a$, $\tau$. 
\end{lemma}

In the sparse case the similar result to Lemma \ref{init}  holds true giving a bound on the error of the initial oracle estimator. 
 \begin{lemma}
 \label{init_sparse}
The error of initial oracle estimation is bounded as follows  
 $r(t_0) \leq \frac{\alpha}{\sqrt{t_0}},$
with probability $1-C_0(n\vee T)^{-2}$, where 
 $$
 \alpha =\left(\frac{1}{ \lambda_d } C_1\lambda_1 h_d M^{1/2}(t_0) + C_2\frac{\lambda_1+1}{\lambda_d}\right)\sqrt{ {\log(n\vee T)}},
 $$  
 where $C_0$ and $C_2$ are constants and $C_1$ depends on $r$.
\end{lemma} 
\begin{lemma}
\label{lemma_7}
Thus after $t$ iterations, $t=t_0+1,\dots,T$, with the probability $1-C_0(n\vee  t)^{-2}$ one has 
\begin{equation*}
\begin{split}
l (V,\widehat{V}^{\circ}(t)) \leq  & C_1 h_d^2 M(t)  \frac{\log(n\vee t)}{t} +  C_2 \frac{\lambda_1+1}{ \lambda_d^2} \frac{\log(n\vee t)}{t},
\end{split} 
\end{equation*}
where $C_1$  depends  on $d$, $r$, $\tau$, $a$,  and $C_2$ depends  on  $\tau$.  
\end{lemma}


The convergence of the oracle scheme  doesn't  immediately imply the convergence of the SCPAST estimators. The following two lemmas state that with the  high probability $\widehat{V}^{\circ}(t)=\widehat{V}(t)$. Thus the bound in Lemma \ref{lemma_7} holds for SCPAST and the Theorem \ref{scpast_theorem} is justified.

 \begin{lemma}
\label{oracle_init}
For $\gamma_0 \geq 3\sqrt{2} \frac{\log(n\vee T)}{\log(n \vee t_0)}$ with probability $1-C_0 (n\vee T)^{-2}$ the initial oracle estimate coincide with the initial SPCA estimate, that is, $\widehat{V}^{\circ}(t_0) = \widehat{V}(t_0)$. 
\end{lemma} 

 \begin{lemma}
 \label{lemma64}
With probability $1-C_0(n\vee t)^{-2}$ for $u=t_0+1,\dots,t$  the oracle SCPAST and SCPAST solutions coincide  $\widehat{V}^{\circ}(t)=\widehat{V}(t)$.
\end{lemma}
\section*{Conclusions}
We developed a new method SCPAST  based on constraint projection approximation subspace tracking method for subspace tracking  in the sparsity assumptions on the underlying signal eigen subspace.   The thesholding  step was introduced in order to ensure the sparsity of the solution. We presented the non-asymptotical  bounds for the errors of subspace recovery with  SCPAST and CPAST as well as the empirical studies of the methods.  
 The results of experiments show that SCPAST method allows to obtain sparse representation of the leading eigenvector 
 of music signals and might be used for  adaptive compression of the musical signal in the spectral domain.

\section*{Appendix A.  Proofs of Lemmas}
\subsection*{Proof of Lemma 1}
From the definition \eqref{tan}  $\tan\phi_{d} (\widehat{\Sigma}(t)\widehat{V}(t-1),V)  =  \max_{\|x\|_{2}=1}\frac{\|\bar{{V}}^{\top}\widehat{\Sigma}(t )\widehat{V}(t-1) x\|}{\|V^{\top}\widehat{\Sigma}(t )\widehat{V}(t-1)x\|}$. Using $\widehat{\Sigma}(t)=\widehat{\Sigma}(t)-\Sigma+\Sigma$ the latter maximum can be bounded by
\begin{align*}
\max_{\|x\|_{2}=1}\frac{  \|\bar{{V}}^{\top}\widehat{V}(t-1)x\|+\|\bar{{V}}^{\top}(\widehat{\Sigma}(t)-\Sigma)\widehat{V}(t-1)x\|}{(\lambda_{d}+1)\|V^{\top}\widehat{V}(t-1 )x\|-\|V^{\top}(\widehat{\Sigma}(t)-\Sigma)\widehat{V}(t-1)x\|}.
\end{align*}
Using \eqref{cos}
\begin{align*}
\max_{\|x\|_{2}=1}&\frac{\|\bar{V}^{\top}(\widehat{\Sigma}(t)-\Sigma)\widehat{V}(t-1)x\|}{\|V^{\top}\widehat{V}(t-1)x\|}  
 \leq\frac{ \|\bar{V}^{\top}(\widehat{\Sigma}(t)-\Sigma)\|}{\cos \phi_{d}(\widehat{V}(t-1),V)},\\
\max_{\|x\|_{2}=1}&\frac{\|V^{\top}(\widehat{\Sigma}(t)-\Sigma)\widehat{V}(t-1)x\|}{\|V^{\top}\widehat{V}(t -1)x\|}
  \leq\frac{\|V^{\top}(\widehat{\Sigma}(t)-\Sigma)\|}{\cos\phi_{d}(\widehat{V}(t-1),V)}.
\end{align*}
From \eqref{cov_decomp}
\begin{equation}
\label{perturb_sigma}
\begin{split}
\widehat{\Sigma}(t) =&
  \Sigma + \frac{1}{t} \Xi(t)\Xi^{\top}(t) -I_n\\ &+\frac{1}{t} V \Lambda^{1/2}_d U^{\top} (t)\Xi^{\top} (t)+ \frac{1}{t}\Xi(t)U(t)\Lambda^{1/2}_dV^{\top}\\& + V \Lambda^{1/2}_d\left(\frac{1}{t} U^{\top} (t)U(t)-I_d\right)\Lambda^{1/2} _dV^{\top}.
\end{split}
\end{equation}
Therefore using $\bar{V}\bar{V}^{\top}+VV^{\top}=I_n$
\begin{align*} 
\| \bar{V}^{\top}(\widehat{\Sigma}(t)-\Sigma)\|  
\leq &   \left\| \frac{1}{t} \bar{V}^{\top}\Xi(t) [\bar{V}^{\top}\Xi(t)]^{\top}-I_{n-d}\right\|
 \\& + \frac{1}{t} \|  \bar{V}^{\top}\Xi(t) [{V}^{\top}\Xi(t)]^{\top} \| \\&+ \sqrt{\lambda_1}\frac{1}{t}\|\bar{V}^{\top}\Xi(t)U(t)\|, \\
 \|{V}^{\top}(\widehat{\Sigma}(t)-\Sigma)\| \leq  &  \lambda_1 \left\|\frac{1}{t} U^{\top} (t)U(t)-I_d\right\|\\&+\left\| \frac{1}{t} {V}^{\top}\Xi(t) [{\bar{V}}^{\top}\Xi(t)]^{\top}\right \|
 \\& + \left\|\frac{1}{t}  {V}^{\top}\Xi(t)\left[{V}^{\top}\Xi(t)\right]^{\top}-I_d\right\|  \\&+ 2 {\sqrt{\lambda_1}}  \left\| \frac{1}{t}\Xi(t)U(t)\right\| . 
\end{align*}
Using $\sqrt{\lambda_1}\vee 1\leq \sqrt{\lambda_1+1}\leq \lambda_1+1$, Lemma    \ref{norm_1}  and Lemma \ref{norm_2} with $p=\sqrt{6}$  we bound the terms to the right of the above two inequalities with the probability $1-C_0 (n \vee t)^{-3}$,   for big enough $t$   
\begin{equation}
\begin{split}
\label{proj_bound}
 \|\bar{V}^{\top}(\widehat{\Sigma}(t)-\Sigma)\|  & \leq \sqrt{\lambda_1+1} E(t),\\ 
 \| V^{\top}(\widehat{\Sigma}(t)-\Sigma)\| & \leq   (\lambda_1 + 1) E(t),
 \end{split}
\end{equation}
where 
$E (t ) =5 \sqrt{\frac{n-d}{t}} + 5\sqrt{6}  \sqrt{\frac{\log(n\vee t)}{t}}$ 
with probability  $1-C_0 (n \vee t)^{-3}$, where $C_0$ is a constant.  
The  statement of the lemma follows from the observation that 
$$
\tan\phi_{d}(\widehat{V}(t),V)  = \tan\phi_{d} (\widehat{\Sigma}(t)\widehat{V}(t-1),V). 
$$ 
 \subsection*{Proof of Lemma 2}
 
 Lemma \ref{lemma_1}   gives a probabilistic bound on  the error of subspace estimation $l (\widehat{V}(t),V)$  based  on the previous iteration $l (\widehat{V}(t-1),V)$.  Or goal is to bound $l (\widehat{V}(t),V)$  for $t\in\left \{t_0+1,\dots,T\right\}$. 
Due to Lemmas \eqref{norm_1} and \eqref{norm_2} we get 
for $p>1$,  $u=t_0+1,\dots,t$ the term $ \sqrt{u} \left \| \frac{1}{u}  {V}^{\top}\Xi(u)\left[{V}^{\top}\Xi(u)\right]^{\top}-I_d \right\|$ is bounded from above by  $3(\sqrt{d}+p\sqrt{\log(n\vee t)})$, $\sqrt{u}  \left \| \frac{1}{u} \bar{V}^{\top}\Xi(u) [\bar{V}^{\top}\Xi(u)]^{\top}-I_{n-d}\right\|$ by $3(\sqrt{n-d}+p\sqrt{\log(n\vee t)})$, $\sqrt{u} \left\|\frac{1}{u} {V}^{\top}\Xi(u) [{\bar{V}}^{\top}\Xi(u)]^{\top} \right\|$ by $\sqrt{1+2p\frac{\log(n\vee t)}{\sqrt{u}}}(\sqrt{n-d}+\sqrt{d}+p\sqrt{\log(n\vee t)})$. Finally, $\sqrt{u}\left\| \frac{1}{u}\bar{V}^{\top}\Xi(u)U(u) \right\|$ is bounded by $$  \sqrt{1+2p\frac{\log(n\vee t)}{\sqrt{u}}}(\sqrt{n-d}+\sqrt{d}+p\sqrt{\log(n\vee t)}).$$
Each of the bounds holds with the probability $1-(n\vee t)^{-3}$ for $p=\sqrt{6}$. Using the union bound we get the statement of the Lemma for the intersection of events  with the probability 
\mbox{$1-C_0 (t-t_0) (n\vee t)^{-3} $}.

\subsection*{Proof of Lemma 3}

The proof is based on Davis $\sin \theta$ Theorem  \ref{Davis}, Lemma \ref{norm_1} (see Appendix B), and Weyl's theorem \cite{stewart1990matrix}. From Davis $\sin\theta$ Theorem
\begin{equation}
\label{v0}
\begin{split}
l(V,\widehat{V}(t_0)) \leq \frac{\left\|(\widehat{\Sigma}(t_0)-\Sigma)V \right\|^2}{\left(\lambda_d+1-\lambda_{d+1}\left(\widehat{\Sigma}(t_0)\right)\right)^2},
\end{split}
\end{equation}
where $\lambda_{d+1}(A)$ is a $(d+1)$-th singular value of the matrix $A^{T}A$.
Weil's theorem gives for $j=1,\dots,n$
 \begin{equation*}
 |\lambda_{j}+1-\lambda_j(\widehat{\Sigma}(t_0))|\leq \|\widehat{\Sigma}(t_0)-\Sigma\|.
 \end{equation*}
Therefore the denominator in \eqref{v0} may be bounded as
$$
|\lambda_d +1 - \lambda_{d+1}(\widehat\Sigma(t_0))|\geq 
\lambda_{d} - 2\| \widehat{\Sigma}(t_0)-\Sigma\|.
$$
From \eqref{perturb_sigma}
 \begin{align*}
 \| \widehat{\Sigma}(t_0)-\Sigma\|\leq & \lambda_1 \left\|\frac{1}{t_0}U(t_0)^{\top} U(t_0)-I_d \right\|\\&+  \left \|\frac{1}{t} \Xi(t_0)\Xi^{\top}(t_0)-I_n\right\| \\&+ 2 \sqrt{\lambda_1}\left\| \frac{1}{t_0}  U^{\top} (t_0)\Xi^{\top} (t_0)\right \|,
\end{align*}
 by  Lemma \ref{norm_1} and \ref{norm_2} 
 $$
 |\lambda_d +1 - \lambda_{d+1}(\widehat\Sigma(t_0))|\geq   (1+o(1))\lambda_d.
 $$
 From \eqref{proj_bound} and Lemma \ref{norm_1} one has that with probability $1-C_0(n\vee t)^{-2}$
$$
\|(\widehat{\Sigma}(t_0)-\Sigma)V\| \leq  ( \lambda_1+1) \left(\sqrt{\frac{n}{t_0}}+  \sqrt{ \frac{\log (n\vee t)}{t_0} }\right).
 $$
 Combining the last two inequalities we get the statement of Lemma.

\subsection*{Proof of Lemma 4}
 
First we prove \eqref{r_0_bound} for  the  pair $(t_{0},r(t_{0}))$ which satisfies 
by induction  for all $k=1,\ldots,K,$ and some $\rho\in(\alpha_0,1)$ 
\begin{equation}
\sqrt{1-\left(r(t_0)+R\left(\frac{\alpha_{1}}{\alpha_{0}}\right)\frac{\rho}{1-\rho}\frac{1}{\sqrt{t_{0}}}\right)^{2}}-\frac{\alpha_{2}R }{\sqrt{t_{0}}}>\frac{\alpha_{0}}{\rho}.
 \label{eq: bound_up}
\end{equation}
We have for $K=1$
\[
r(t_{0}+1)\leq\frac{\alpha_{0}r(t_{0})+\alpha_{1}\frac{R}{\sqrt{t_{0}+1}}}{\sqrt{1-r^{2}(t_{0})}-\alpha_{2}\frac{R}{\sqrt{t_{0}+1}}}\leq\rho r(t_{0})+\frac{\rho\alpha_{1}}{\alpha_{0}}\frac{R}{\sqrt{t_{0}+1}}.
\]
Furthermore suppose that (\ref{eq: bound_up}) holds for $K=L,$ then $$r^{2}(t+L)\leq \left(\rho^{L}r(t_{0})+R\left(\frac{\alpha_{1}}{\alpha_{0}}\right)\sum_{k=1}^{L}\frac{\rho^{L+1-k}}{\sqrt{t_{0}+k}}\right)^{2}$$ and 
\begin{align*}
r(t_{0}+L+1) & \leq\frac{\alpha_{0}r(t+L)+\alpha_{1}\frac{R}{\sqrt{t+L+1}}}{\sqrt{1-r^{2}(t+L)}-\alpha_{2}\frac{R}{\sqrt{t+L+1}}} \\
 & \leq\rho^{L}r(t_{0})+R\left(\frac{\alpha_{1}}{\alpha_{0}}\right)\sum_{k=1}^{L+1}\frac{\rho^{1+L-k}}{\sqrt{t_{0}+k}}.
\end{align*} 
A sufficient condition for the above formula to hold reads as 
\begin{equation*}
\sqrt{1-\left(\rho^{k-1}r(t_{0})+ \frac{\alpha_{1}R}{\alpha_{0}} \sum_{j=1}^{k-1}\frac{\rho^{k-j}}{\sqrt{t_{0}+j}}\right)^{2}}-\frac{\alpha_{2} R}{\sqrt{t_{0}+k}}>\frac{\alpha_{0}}{\rho}.
\end{equation*}
Note that
$
\sum_{j=1}^{k-1}\frac{\rho^{k-j}}{\sqrt{t_{0}+j}}\leq\frac{\rho}{1-\rho}\frac{1}{\sqrt{t_{0}}},
$
therefore the above condition  is fulfilled given \eqref{eq: bound_up}. Furthermore 
\begin{align*}
r(t_{0}+K+1)\leq\rho^{K}r(t_{0})+R\left(\frac{\alpha_{1}}{\alpha_{0}}\right)\sum_{k=1}^{K+1}\frac{\rho^{1+K-k}}{\sqrt{t_{0}+k}},
\end{align*}
where for $K>K_{0}(\rho)$, $t_{0}>1$ and $j_{K,\rho}=\log(K)/(2\log(1/\rho))$
\begin{align*}
\sum_{k=1}^{K}\frac{\rho^{ K-k}}{\sqrt{t_{0}+k}}  \leq &  \sum_{j=0}^{j_{K,\rho}}\frac{\rho^{j}}{\sqrt{t_{0}+K-j}} + \sum_{j=j_{K,\rho}+1}^{K-1}\frac{\rho^{j}}{\sqrt{t_{0}+K-j}}\\
 & \leq\frac{1}{1-\rho}\frac{1}{\sqrt{t_{0}+K-j_{K,\rho}}}+\frac{1}{1-\rho}\frac{1}{\sqrt{K+t_{0}}}
\end{align*}
and 
\[
r(t_{0}+K+1)\lesssim \frac{\alpha_{1}}{\alpha_{0}}\frac{\rho}{1-\rho} \frac{R}{\sqrt{K+t_{0}+1}}.
\]
From  \eqref{eq: bound_up} the  condition on  the starting value $r(t_0)$ is  
\begin{equation}
\label{r0_condition1}
 r(t_0) \leq \sqrt{1-\left(\alpha_{2}\frac{R}{\sqrt{t_{0}}}+\frac{\alpha_{0}}{\rho}\right)^2} - \frac{R\alpha_{1}}{\alpha_{0}} \frac{\rho}{1-\rho}\frac{1}{\sqrt{t_{0}}} .
\end{equation}
Thus   the number of initial observations $t_0$ for \eqref{r0_condition1}
 to be satisfied given 
  $r(t_0)\leq \alpha\frac{1}{\sqrt{t_0}}$ reads as
$$ 
  \alpha\frac{1}{\sqrt{t_0}} \leq  \sqrt{1-\left(\alpha_{2}\frac{R}{\sqrt{t_{0}}}+\frac{\alpha_{0}}{\rho}\right)^2} -R\left(\frac{\alpha_{1}}{\alpha_{0}}\right)\frac{1}{1-\rho}\frac{1}{\sqrt{t_{0}}}.
 $$
Therefore, taking into account \eqref{alphas} and $\rho>\alpha_0$, 
the sufficient condition on $\sqrt{t_0}$ is 
$$ 
 \sqrt{t_0} \geq 
\frac{2 \alpha_{2} {R}}{ \left(1-\frac{\alpha^2_{0}}{\rho^2}\right)  }+ \frac{2 R\left(\frac{\alpha_{1}}{\alpha_{0}}\right)\frac{\rho}{1-\rho}}{\sqrt{ 1-\frac{\alpha^2_{0}}{\rho^2}  }}.
 $$ 
From Lemma \ref{init} and  \eqref{alphas}
${\alpha} =  {R_{\max}}  \frac{\alpha_2}{1-\alpha_0}.
$
 Set $\rho=\rho(\epsilon) = 1-\epsilon(1- \alpha_0)$. It is easy to check that  $\alpha_0<\rho(\epsilon)<1$ for $\epsilon \in (0,1/2]$. Recall $R_{\max}=R(T)$, therefore 
$$
 \sqrt{t_0} \geq    \frac{2 R_{\max} \alpha_{2}}{\epsilon(1-\alpha_0)^{3/2}}  \frac{1}{\sqrt{ 1-\epsilon }}.
 $$ 
The value $K_{0}(\rho)$ might be defined by 
$ 
\rho^{K} \leq \frac{\alpha_1}{\alpha_0}\frac{R}{\sqrt{t_0+K}},
$ 
thus (using $|\ln (1-x)|\geq x$ for $x\in(0,1)$) it is sufficient to set 
$ 
K\geq \frac{1}{\epsilon(1-\alpha_0)}\ln \left(\frac{\alpha_0}{\alpha_1}\sqrt{T}\right).
$ 
Put $\epsilon = 1/2$ to get the result. 

\subsection*{Proof of Lemma 5}
Using the triangle inequality
\begin{align}
\label{triangsin}
\sqrt{l (\widehat{V}^{\circ}(t), V)}\leq\sqrt{ l (\widehat{\Upsilon}^{\circ}(t) , V)} +  \sqrt{l( \widehat{\Upsilon}^{\circ}(t),\widehat{V}^{\circ}(t))}.
\end{align}
We bound the first term as 
$$
l^{1/2} (\widehat{\Upsilon}^{\circ}(t) , V)\leq \tan\phi_d(\widehat{\Sigma}^{\circ}(t)\widehat{V}^{\circ} (t-1),V). 
$$Using the variational definition of $\tan\phi_d$
\begin{equation*}
\label{itersparse}
\begin{split}
 &l^{1/2} (\widehat{\Upsilon}^{\circ}(t) , V)\leq  \max_{\|x\|_{2}=1}\frac{\|\bar{{V} }^{\top}\widehat{\Sigma}^{\circ}(t)\widehat{V}^{\circ} (t-1) x\|}{\|V^{  \top}\widehat{\Sigma}^{\circ}(t)\widehat{V}^{\circ} (t-1) x\|}  
\end{split}
\end{equation*}
The right hand side may be bounded with 
$$
 \max_{\|x\|_{2}=1}\frac{\|\bar{V}^{  \top} \Sigma \widehat{V}^{\circ}(t-1)x\| + \|\bar{V}^{  \top} ( \widehat{\Sigma}^{\circ}(t)-\Sigma ) \widehat{V}^{\circ}(t-1)x\|}{\|V^{ \top} \Sigma\widehat{V}^{\circ}(t-1)x\| - \|V^{ \top} (\widehat{\Sigma}^{\circ}(t)-\Sigma) \widehat{V}^{\circ}(t-1)x\|}.
$$
Triangle inequality gives 
\begin{align*}
  \|(\widehat{\Sigma}^{\circ}(t)   -  {\Sigma} ) \bar{V}  \|  \leq &\|(\widehat{\Sigma}^{\circ}(t)-\Sigma^{\circ}(t))\bar{V}\|\ +  \| ({\Sigma}^{\circ}(t)   -  {\Sigma} ) \bar{V}  \| . 
\end{align*}
Note that 
 \begin{align}
\label{mat_break1}
& \widehat{\Sigma}^{\circ}(t)-\Sigma^{\circ}(t)   =
 \left[\begin{matrix}
 \widehat{\Sigma}_{S}(t)-\Sigma_{S}(t) & 0\\
 0 &0
  \end{matrix}\right]
,\\ \label{mat_break2}
&{\Sigma}^{\circ}(t)   -  {\Sigma}  =
\left[\begin{matrix}
0 & - V_{S}(t) \Lambda_d V_{N}^{\top}(t) \\
- V_{N}(t) \Lambda_d V_{S}^{\top} (t) & -V_{N}(t) \Lambda_d V_{N}^{\top}(t)
 \end{matrix}
 \right],
 \end{align}
where  $V_{S}(t)$ is a submatrix of $V$ with the row indices in $S(t)$. 
Decompose  $\widehat{\Sigma}_{S}(t)-{\Sigma}_{S}(t)$ using \eqref{perturb_sigma} and \eqref{mat_break1}
\begin{equation*}
\begin{split}
&\|(\widehat{\Sigma}^{\circ}(t)-\Sigma^{\circ}(t))\bar{V}\| \leq  \lambda_1 \|\bar{V}_{S}^{\top} V_{S}(t) \|  \left\|  \frac{1}{t}U(t)^{\top}U(t)-I_d  \right\|\\ 
  & + \left\|\frac{1}{t}\Xi_{S}(t)\Xi_{S}^{\top}(t)-I_{S}(t) \right\| 
 + 2  \sqrt{\lambda_1}\left\|\frac{1}{t}  U(t)^{\top} \Xi^{\top}_{S}(t) \right\|,
\end{split}
 \end{equation*}
  where $\Xi_{S}(t)$ is $t \times {\rm  card }(S(t))$ matrix, $U(t)$ is $t\times d$ matrix. The elements of both matrices are  i.i.d. $\mathcal{N}(0,1)$.  
 Using $\bar{V}^{\top} V=\bar{V}_{S}^{\top}(t) V_{S}(t)+\bar{V}_{N}^{\top}(t) V_{N}(t)=0$ we may bound  
 $$
 \|\bar{V}_{S}^{\top} V_{S}(t) \| \leq \|\bar{V}_{N}^{\top} V_{N}(t) \|\leq \|V_{N}(t)\|\leq \|V_{N}(t)\|_{\rm F},
 $$
 where  $\|\cdot\|_{\rm F}$ is Frobenius norm, i.e. $\|A\|_{\rm F}= \sqrt{{\rm tr}(A^{\top}A)}$  for any  matrix $A$,  is small since it depends only on the components of the eigenvectors below the corresponding thresholds (see Lemma \ref{A1} and  definition \eqref{M})
\begin{equation*}
\begin{split}
 \|V_{N}(t)\|^2_{\rm F}= &\sum_{i=1}^d \|v_{i,N}(t)\|^2 
 \\
\leq & \sum_{j=1}^d\left[\frac{2}{2-r}   \frac{ {t}^{\frac{r}{2}} s_j^r/(bh_j)^r }{[\log(n\vee t)]^{\frac{r}{2}}}  \wedge  n\right] b^2 h_j^2   \frac{\log(n\vee t)}{t}  \\
 \leq &  C M (t) h_d^2  \frac{\log(n\vee t)}{t},
\end{split}
\end{equation*}
where $C$ depends on $d,r$. 

From Lemma \ref{norm_1}, \ref{norm_2}  and \ref{sparse78} (see Appendix B) 
 with the probability $1-C_0 (n\vee t)^{-3}$ 
  one can bound   
\begin{equation*}
\begin{split}
 \|(\widehat{\Sigma}^{\circ}(t)-\Sigma^{\circ}(t))\bar{V}\|  
\leq &C_1\lambda_1 h_d \frac{M^{1/2}(t)}{\sqrt{t}}\\& + C_2(\sqrt{\lambda_1}\vee 1)\frac{\sqrt{ \log(n\vee t)}}{\sqrt{t}} .
\end{split} 
 \end{equation*}
From \eqref{mat_break2} 
$
\|\bar{V}^{\top}(\Sigma^{\circ}(t)-\Sigma)\| \leq \lambda_1 \|V_{N}(t)\|.
$
Thus
\begin{equation}
 \label{E1circ0}
\begin{split}
 \|(\widehat{\Sigma}^{\circ}(t)-\Sigma^{\circ}(t))\bar{V}\|  
\leq & C_1\lambda_1 h_d M^{1/2}(t)\sqrt{\frac{\log(n\vee t)}{t}} \\
&+ C_2(\sqrt{\lambda_1}\vee 1)\sqrt{ \frac{\log(n\vee t)}{t} }.
\end{split}
\end{equation}
 where $C_1$ depends on $r$, $d$  and $C_2$ is a constant.
Similarly, from \eqref{mat_break2}   $
\|V^{\top}(\Sigma^{\circ}(t)-\Sigma)\|\leq |V_{N}(t)\|(1+o(1)) 
$ 
 and 
 \begin{equation}
 \label{E2circ0}
 \begin{split}
 \|{V}^{\top}(\widehat{\Sigma}^{\circ}(t)-\Sigma)\|\leq &  
  C_1\lambda_1 h_d M^{1/2}(t)\sqrt{\frac{\log(n\vee t)}{t}} \\&+ C_2(\sqrt{\lambda_1}\vee 1)\sqrt{ \frac{\log(n\vee t)}{t} },
 \end{split}
\end{equation}
 where $C_1$ depends on $r$, $d$  and $C_2$ is a constant.
 
The  bound on   $l(\widehat{\Upsilon}^{\circ}(t),\widehat{V}^{\circ}(t)) =l(\widehat{\Upsilon}^{\circ}(t),\widehat{\Upsilon}^{\circ,\beta}(t))$ relies on Wedin's $\sin \theta$ \mbox{Theorem \ref{Wedin}} (see Appendix B)
\begin{equation}
\label{Wedin_or}
  l(\widehat{\Upsilon}^{\circ}(t),\widehat{\Upsilon}^{\circ,\beta}(t))\leq \frac{\|\widehat{\Sigma}^{\circ}(t)\widehat{V}^{\circ} (t-1)-\widehat{\Sigma}^{\circ,\beta}(t) \|^2 }{\lambda_{d}(\widehat{\Sigma}^{\circ}(t)\widehat{V}^{\circ} (t-1))}.
\end{equation}
Note that 
$
\|\widehat{\Sigma}^{\circ}(t)\widehat{V}^{\circ} (t-1) - \widehat{\Upsilon}^{\circ,\beta}(t-1) \| \leq  \|Z(t) \|_F,
$
where  $Z_{ij}(t)$ is a matrix with the entries 
$Z_{ij}(t)= \beta_{j}(t)$ if $i\in S(t)$ and $Z_{ij}(t)= 0$ if $i\in N(t)$.
Thus \mbox{$\|\widehat{\Sigma}^{\circ}(t)\widehat{V}^{\circ} (t-1)-\widehat{\Upsilon}^{\circ,\beta}(t) \|^2  \leq    CM(t)\sum_{i=1}^d\beta_{i}^2(t) $}   and 
from \eqref{beta_def}   
$$
\sum_{i=1}^d\beta_{i}^2(t)  
 \leq      a^2 \frac{\log(n\vee t)}{t} \sum_{i=1}^d (\lambda_i+1)  
  \leq    d a^2   \lambda_d^2  \frac{\log(n\vee t)}{t}h_d^2   
$$
That is 
$$\|\widehat{\Sigma}^{\circ}(t)\widehat{V}^{\circ} (t-1)-\widehat{\Upsilon}^{\circ,\beta}(t) \|^2  \leq    CM(t) \lambda_d^2  \frac{\log(n\vee t)}{t}h_d^2, $$ 
where $C^{\prime} $ depends on $d$, $a$ and $r$.

To bound the denominator of \eqref{Wedin_or}  note that one may decompose  $\|\Sigma\widehat{V}^{\circ}(t-1)x\|^2= 
\|\Sigma z_1\|^2+\|\Sigma z_2\|^2,$ where  $\widehat{V}^{\circ}(t-1)x = z_1 + z_2$ and $z_1\in{\rm ran}(V)$ and $z_2\in {\rm ran}(\bar{V})$.
Thus $\|\Sigma\widehat{V}^{\circ}(t-1)x\|^2\geq \|\Sigma z_1\|^2$.  Using $z_1\in {\rm ran}(V)$  one has  $$\|\Sigma z_1\|\geq (\lambda_d+1)\|z_1\|\geq (\lambda_d+1)\cos(V,\widehat{V}^{\circ}(t-1))$$ and taking into account \eqref{mat_break2} we get 
\begin{equation*}
\begin{split}
{\lambda^{1/2}_{d}\left(\widehat{\Sigma}^{\circ}(t)\widehat{V}^{\circ} (t-1)\right)}
 \geq & ({\lambda}_{d}+1)   \cos(V,\widehat{V}^{\circ}(t-1))\\&- \|\widehat{\Sigma}^{\circ}(t)-\Sigma^{\circ}(t)\|  - \lambda_1 \|V_{N}(t)\|. 
\end{split}
\end{equation*}
Thus using \eqref{mat_break1} and  Lemmas \ref{norm_1}, \ref{norm_2} and summarizing the bounds for denominator and nominator in \eqref{Wedin_or} we get 
$$
l^{1/2}(\widehat{\Upsilon}^{\circ}(t),\widehat{V}^{\circ}(t))\leq \frac{  \lambda_d   C {M^{1/2}(t)}\sqrt{ \frac{\log(n\vee t)}{t}}h_d  }{ ({\lambda}_{d}+1) \cos(V,\widehat{V}^{\circ}(t-1)) -   E^{\circ}(t)} ,
$$ 
where 
$
 E^{\circ}(t) =\left (C_1\lambda_1 h_d M^{1/2}(t) + C_2(\sqrt{\lambda_1}\vee 1)\right)\sqrt{ \frac{\log(n\vee t)}{t} }.
$

Combining  the above inequality, \eqref{E1circ0}, \eqref{E2circ0},   \eqref{triangsin}   and the spectral gap condition \eqref{AD} we get the result in the flavour of \eqref{lemma_1}, that is with probability $1-C_0 (n\vee t)^{-3}$ for one step of SCPAST algorithm.
To get the bounds for $u=t_0+1,\dots,t$ simultaneously, similarly to Lemma \ref{lemma_2} define the events, each of which occurs with probability $1-C_0 (n\vee t)^{-3}$, namely that $\sqrt{u}  \left\| \frac{1}{u}U(u)^{\top}U(u)-I_d\right\|$ is bounded from above by $  2(\sqrt{d}+p\sqrt{\log(n\vee t)})$, $ \sqrt{u}  \left\|\frac{1}{u}\Xi_{S}(u)\Xi_{S}^{\top}(u)-I_{S}(u) \right\|$ by $2(\sqrt{{\rm card}(S(t)})+p\sqrt{\log(n\vee t)})$,  $\sqrt{u} \left\| \frac{1}{u}  U(u)^{\top} \Xi^{\top}_{S}(u) \right\|$ by  $ \sqrt{1+2p\frac{\log(n\vee t)}{\sqrt{u}}} (\sqrt{{\rm card}(S(t)})+  \sqrt{d}+p\sqrt{\log(n\vee t)} ).$ 
 Taking the intersection of the above events for $u=t_0+1,\dots,t$ and using Lemma \ref{sparse78}, we get the statement of the Lemma.
 
\subsection*{Proof of Lemma 6} 
Using Wedin $\sin \theta$ Theorem \ref{Wedin} (Appendix B)
\begin{equation}
\begin{split}
l(V,\widehat{V}^{\circ}(t_0)) \leq \frac{\|V ^{\top}(\Sigma-\widehat{\Sigma}^{\circ}(t_0))\|^2 }{(\lambda_{d}- {\lambda}_{d+1}(\widehat{\Sigma}^{\circ}(t_0)) ^2 }. 
\end{split}
\end{equation}
Using Weyl theorem \cite{stewart1990matrix} it may be shown that $\lambda_{d+1}(\widehat{\Sigma}^{\circ}(t_0)) = \lambda_{d+1} +o (\lambda_{1})$ and thus 
$ 
|\lambda_{d}-\lambda_{d+1}(\widehat{\Sigma}^{\circ}(t_0))| \geq  \lambda_d (1+o(1)).
$ 
From  \eqref{E2circ0} with probability $1-(n\vee T)^{-2}$
\begin{equation*} 
 \begin{split}
 \|(\widehat{\Sigma}^{\circ}(t_0)-\Sigma){V}\|\leq &  
C_1\lambda_1 h_d M^{1/2}(t_0)\sqrt{ \frac{\log(n\vee t)}{t_0} } \\&+ C_2(\sqrt{\lambda_1}\vee 1) \sqrt{ \frac{\log(n\vee T)}{t_0}}. 
  \end{split}
\end{equation*}
Thus 
 $r(t_0) \leq \frac{\alpha}{\sqrt{t_0}}$
 holds with probability $1-(n\vee T)^{-2}$, where 
 $$
 \alpha =\left(\frac{1}{ \lambda_d } C_1\lambda_1 h_d M^{1/2}(t_0) + C_2\frac{\sqrt{\lambda_1+1}}{ \lambda_d }\right)\sqrt{ {\log(n\vee T)}}  . 
 $$
  
\subsection*{Proof of Lemma 7}
The proof  follows from Lemma \ref{lemma_3} applied to \eqref{iter_alg_0s}
with 
$
\alpha_0= \frac{ 1}{\lambda _d +1}$, $\alpha_1 =\alpha_2 = \frac{\lambda_1 +1}{\lambda_d+1}, 
$
and initial conditions given by Lemma \ref{init_sparse}.
 \subsection*{Proof of Lemma 8} 
Following \cite{paul2012augmented} define
$
\eta_j = \sum_{i=1}^d \lambda_i v^2_{ji}
$
, $j=1,\dots,n$
and  for $0<a_{-}<1 $  
define
$ 
G^{+}=\left\{j: \eta_j>a_{-}\gamma_0 \sqrt{\frac{\log(n\vee t_0)}{t_0}} \right\}.
$ 
To show that $\widehat{V}^{\circ}(t_0)=\widehat{V}(t_0)$ one has to prove that for the proper choice of $\gamma_0$ and $a_{-}$ it holds 
$G \subseteq G_{+} \subseteq  S(t_0)$ with probability $1-C_0 (n\vee T)^{-2}$. To show that we first note that  $\widehat{\Sigma}_{jj}(t_0) \sim  (1+\sum_{i=1}^d\lambda_i v^2_{ji})\xi/t_0$, where $\xi$ is $\chi^2_{t_0}$ r.\,v.  Therefore
\begin{equation*}
\begin{split}
\mathbf{P}(G\not\subset G^{+}) =  &\mathbf{P}\left\{ \bigcup\limits_{j\not\in G^{+}} \left(\widehat{\Sigma}_{jj}(t_0) > 1+\gamma_0\sqrt{\frac{\log (n\vee t_0)}{t_0}} \right) \right\} \\
 \leq & \sum_{j\not\in G^{+}} \mathbf{P}\left\{\widehat{\Sigma}_{jj} (t_0)> 1+\gamma_0 \sqrt{\frac{\log (n\vee t_0)}{t_0}} \right\} \\
    \leq & n   \mathbf{P}\left\{\frac{\xi}{t_0}-1 >  \frac{\gamma_0  (1-a_{-})  \sqrt{\frac {\log (n\vee t_0)} {t_0}}}{  1+ a_{-} \gamma_0 \sqrt{\frac{\log (n\vee t_0)}{t_0}} } \right\} \\
    \leq & \frac{\sqrt{2}n}{\gamma_0}\exp\left\{ -\frac{\gamma_0^2 (1-a_{-})^2  \log (n\vee t_0) }{4\left(1+ a_{-} \gamma_0 \sqrt{\frac{\log (n\vee t_0)}{t_0}}\right)^2} \right\}  \\
    \leq & \frac{\sqrt{2}}{\gamma_0} n  (n\vee t_0)^ {  - (\gamma_0^2 (1-a_{-})^2/ 4)(1+o(1))}. 
\end{split}
\end{equation*}
Thus $ G \subset G^{+}$ 
holds with probability $1-C_0 (n\vee T)^{-2}$, e.g. for $a_{-}=1-\sqrt{2}/\sqrt{3}$, $\gamma_0\geq 3\sqrt{2}\sqrt{\frac{\log(n\vee T)}{\log(n \vee t_0)}}$. 
Note that for any $j \in G^{+}$ there exists $i\in \{1,\dots,d\}$, $\lambda_i v_{ji}^2\geq \frac{a_{-}\gamma_{0}}{d}\sqrt{\frac{\log (n\vee t_0)}{t_0}}$, thus for $G^{+}\subset S(t_0)$  to hold it is sufficient that 
$$
\frac{a_{-}}{d\lambda_i}\sqrt{\frac{\log(n\vee T)}{t_0}} > b \frac{\lambda_i+1}{\lambda_i^2} \frac{\log (n\vee t_0)}{t_0}.
$$
Thus for sufficiently big $T$, 
  $G\cap S(t_0)=G$, that is  $\widehat{V}^{\circ}(t_0)=\widehat{V}(t_0)$ with probability $1-C_0(n\vee T)^{-2}$.

\subsection*{Proof of Lemma 9}

From Lemma  \ref{oracle_init} with probability $1-(n\vee T)^{-2}$ the results of the original and oracle version of the zero-step estimation procedure coincide, that is $\widehat{V}^{\circ}(t_0)=\widehat{V}(t_0)$. First let us show that the similar statement holds for 
 $\widehat{V}^{\circ}(t_0+1)$ and $\widehat{V}(t_0+1)$. Denote $t_1=t_0+1$.
 On the event for which $\widehat{V}^{\circ}(t_0)=\widehat{V}(t_0)$ holds it is true that
 $\widehat{\Upsilon}(t_1)  = \widehat{\Sigma}(t_1) \widehat{V}(t_0) = \widehat{\Sigma}(t_1) \widehat{V}^{\circ}(t_0).$   
From the construction of $\widehat{V}^{\circ}(t_0)$, the submatrix $\widehat{V}^{\circ}_{N}(t_0)$  has zero entries. Note that $S(t_0)\subseteq S(t_1)$ and $N(t_1)\subseteq N(t_0)$. Thus 
 \begin{equation}
 \label{Ups_dec}
  \widehat{\Upsilon}(t_1)_{k,l}= \widehat{\Sigma}_{k,S}(t_1) \widehat{v}^{\circ}_{l,S}(t_0),
\end{equation}  
 where $\widehat{v}^{\circ}_{l,S}(t_0)$ is a vector of size ${\rm card} (S(t_1))$ containing the components of  $\widehat{v}^{\circ}_{l}(t_0)$ indexed by ${S(t_1)}$, $\widehat{\Sigma}_{k,S}(t_1) $ is a row containing the components of $k$-th row of  $\widehat{\Sigma}(t_1)$ indexed by ${S(t_0)}$.
 Let us show that for $k\in N(t_1)$ with high probability, which is equivalent to  
 $$
  \widehat{\Upsilon}(t_1)_{k,l} \leq  \beta_{l}(t_1) = a \sqrt{(\lambda_l+1) \frac{\log( n\vee t_1)}{t_1}},
 $$
that is during the thresholding step the components from $N(t_1)$ would be set to zero with high probability.  From \eqref{data_repres}
\begin{equation}
\begin{split}
\label{cov_dec_9}
t_1\widehat{\Sigma}_{k,S}(t_1)  =&  V_{k}\Lambda_d^{1/2}  U(t_1)^{\top}U(t_1) \Lambda_d^{1/2}V_{S}^{\top} (t_1)\\&+  \Xi_{k}(t_1) \Xi_{S}^{\top}  (t_1)
\\&+  V_{k}\Lambda_d^{1/2}U(t_1)^{\top}\Xi_{S}^{\top}(t_1) \\
&+    \Xi_{k}(t_1) U(t_1)\Lambda_d^{1/2}V_{S}^{\top}(t_1), 
 \end{split}
\end{equation} 
where $\Xi_{k}(t_1)$ is $k$-th row of $\Xi(t_1)$. 
Denote by ${V}^{\circ}_{S}(t_1)$ a matrix containing the first $d$ eigenvalues of $\Sigma_{S}(t_1)$  as columns (recall \eqref{mat_break}) and  by $\bar{V}^{\circ}_{S}(t_1)$ a matrix with  ${\rm card}(S(t_1))-d$ columns which complete columns of $V^{\circ}_{S}(t_1)$ to the orthonormal basis in $\mathbb{R}^{{\rm card}(S(t_1))}$. Note that 
$$\bar{V}^{\circ}_{S}(t_0) \bar{V}^{\circ,\top}_{S}(t_0) +{V}^{\circ}_{S}(t_0) {V}^{\circ,\top}_{S}(t_0)  = I_{S}(t_0).$$
Plugging in above equality  in \eqref{Ups_dec} (before $V_{S}^{\top}(t_1)$) in the view of \eqref{cov_dec_9} one gets 
\begin{equation*}
 \widehat{\Upsilon}(t_1)_{k,l} = q_{11}+q_{12}+q_{12}+q_{14}+ q_{21}+q_{22}+q_{22}+q_{24},
\end{equation*}
where $q$-s with the  listed below with the first index $1$ depend on $\bar{V}^{\circ}_{S}(t_0)$ and with the first index $2$ depend on ${V}^{\circ}_{S}(t_0)$.
Let us first bound the terms $q_{11}$ and $q_{21}$. To this end we add and subtract $V_{k}\Lambda_d V_{S}^{\top} (t_1)$ in the first term in \eqref{cov_dec_9} and use that $\left\|  \frac{1}{t_1}U(t_1)^{\top}U(t_1)-I\right\|  = o(1)$, thus 
\begin{equation}
\label{q11}
\begin{split}
&|q_{11}|    
 \leq  (1+o(1)) \| V_{k}\Lambda_d  \| \| \bar{V}^{\circ,\top}_{S}(t_0)   \widehat{v}^{\circ}_{l,S}(t_0)\| 
\end{split}
\end{equation}
where it was also used that $\| V_{S}^{\top} (t_1) \bar{V}^{\circ}_{S}(t_0)\| \leq 1$.

Consider $k\in N(t_1)$, that is, $|V_{kl}|=|v_{kl}|\leq b\sqrt{h_i^2\frac{\log (n\vee t_1)}{t_1}}$. Using the definition of $\beta_k(t_1)$ (recall \eqref{beta_def})
\begin{align}\label{rowl}
\| V_k \Lambda_d\|& \leq 
 \frac{b}{a} \beta_k(t_1)  \sqrt{\sum_{i=1}^{d}     \frac{\lambda_i+1}{\lambda_k+1}},\\
\label{rowl12} 
 \| V_k \Lambda^{1/2}_d\|& \leq 
 \frac{b}{a} \beta_k(t_1)  \sqrt{\sum_{i=1}^{d}     \frac{\lambda_i+1}{(\lambda_k+1)\lambda_i}}.
\end{align}
 Thus using \eqref{rowl}  and \eqref{rowl12} the term \eqref{q11} may be bounded as  
$$
|q_{11}|  \leq (1+ o(1))\frac{b}{a} \beta_k(t_1)  \| \bar{V}^{\circ,\top}_{S}(t_0)   \widehat{v}^{\circ}_{l,S}(t_0)\|\sqrt{\sum_{i=1}^{d}     \frac{\lambda_i+1}{\lambda_l+1}}   
$$
and in the same way it can be shown that 
$$
|q_{21} | \leq (1+ o(1))\frac{b}{a} \beta_k(t_1)  \| {V}^{\circ,\top}_{S}(t_0)   \widehat{v}^{\circ}_{l,S}(t_0)\|\sqrt{\sum_{i=1}^{d}     \frac{\lambda_i+1}{\lambda_l+1}}.  
$$
Next 
\begin{equation*}
\begin{split}
|q_{ 12}|  = & \frac{1}{t_1}|\Xi_{k}(t_1) \Xi_{S}^{\top}(t_1)    \bar{V}^{\circ}_{S}(t_0) \bar{V}^{\circ,\top}_{S}(t_0)  \widehat{v}^{\circ}_{l,S}(t_0) | \\
\leq &\frac{1}{t_1} \zeta(k,S(t_1))
\|\Xi_{S}^{\top}(t_1)\|   \|  {V}^{\circ,\top}_{S}(t_0)  \widehat{v}^{\circ}_{l,S}(t_0) \|,
\end{split}
\end{equation*}
where  
$$
\zeta(k,l,S(t_1))= \frac{t_1 q_{12}}{\|\Xi_{S}^{\top} (t_1)   \bar{V}^{\circ}_{S}(t_0) \bar{V}^{\circ,\top}_{S}(t_0)  \widehat{v}^{\circ}_{l,S}(t_0) \|}.
$$
Note that $\widehat{v}^{\circ}_{l,S}(t_0)$ is the $l$-th eigenvector of $\widehat\Sigma_S (t_0)$  and doesn't depend on $\Xi_k(t_1)$, $k\in N(t_1)$, and since $N(t_1)\subseteq N(t_0)$,  $\Xi_k(t_1)$ is independent from $\Xi_S (t_1)$, thus   $\zeta(k,l,S(t_1))$ 
 has $\mathcal{N}(0,1)$ distribution. 
Define the events 
$$
|\zeta(k,l,S(t_1))|\leq  \sqrt{c_1 \log (n \vee t)}
$$
and 
$$
\|\Xi_{S}(t_1)\| \leq \sqrt{t_1}+\sqrt{{\rm card}(S(t_1))}+2 \sqrt{\log (n \vee t)}.
$$
For big enough $t_1$ (guarantied by  \eqref{t0_sparse}) ${t_1}$ dominates ${{\rm card}(S(t_1))}$ and ${\log (n \vee t)}$. Thus 
\begin{equation*}
\begin{split}
|q_{12} |\leq & \frac{1}{t_1} |\zeta(k,S(t_1))|  \|\Xi_{S}^{\top}(t_1)\|   \|  {V}^{\circ,\top}_{S}(t_1)  \widehat{v}^{\circ}_{l,S}(t_0) \|\\
\leq &\frac{1}{a} \left(\frac{c_1}{\lambda_l+1} \right)^{1/2} \beta_l(t_1) \frac{\log(n\vee t)}{\log(n\vee t_1)} \|  {V}^{\circ,\top}_{S}(t_1)  \widehat{v}^{\circ}_{l,S}(t_0) \|. 
\end{split}
\end{equation*}
On the events defined in the end of the proof of Lemma \ref{lemma_5} the bound for  the term $q_{13}$ is as follows 
\begin{equation*}
\begin{split}
|q_{13}| & \leq  \frac{1}{t_1} \| V_{k}\Lambda_d^{1/2}\| \| U(t_1)^{\top}\Xi_{S}^{\top} (t_1) \| \bar{V}^{\circ,\top}_{S}(t_1)    \widehat{v}^{\circ}_{l,S}(t_0) \|\\
&\leq  \frac{b}{a} \frac{{\rm card}(S(t_1))}{\sqrt{\lambda_d}t_1} \frac{\log(n\vee t)}{\log (n \vee t_1)} \beta_l(t_1)   \| \bar{V}^{\circ,\top}_{S}(t_1)    \widehat{v}^{\circ}_{l,S}(t_0) \|.
\end{split}
\end{equation*}
From  $\frac{1}{\lambda_d}\frac{{\rm card}(S(t_1)))}{t_1}=o(1)$  (see Supplementary materials for  \cite{ma2013sparse} p.16) it follows that  $|q_{13}| = o(\beta_l(t_1))$.
To bound the term $q_{14}$ one may utilize the same argument as for $q_{12}$ 
\begin{equation*}
\begin{split}
|q_{14}| \leq  \frac{1}{t_1}|g(k,S(t_1))| \| U(t_1)\Lambda_d^{1/2}\|\| \bar{V}^{\circ,\top}_{S}(t_1)   \widehat{v}^{\circ}_{l,S}(t_0)\|,
\end{split}
\end{equation*} 
where
$$
g(k,l,S(t_1)) = \frac{t_1q_{14}}{\| U(t_1)\Lambda_d^{1/2}V_{S}^{\top}(t_1)  \bar{V}^{\circ}_{S}(t_1) \bar{V}^{\circ,\top}_{S}(t_1) \widehat{v}^{\circ}_{l,S}(t_0)\|}
$$
is independent from $U(t_1)$ and   $\widehat{v}^{\circ}_{l,S}(t_0)$,   furthermore $g(k,l,S(t_1))$ has  $\mathcal{N}(0,1)$ distribution.  
Define the following two events
$
\{\|U(t_1)\| \leq \sqrt{t_1}+\sqrt{d}+2 \sqrt{\log (n \vee t)}\} 
$
and  
$
\{|g(k,l,S(t_1))| \leq \sqrt{ c_1 \log (n \vee t)}\}.
$
On these events  with probability $1-C_0 (n\vee t)^{-3}$
\begin{equation*}
\begin{split}
|q_{14}| \leq &\frac{1}{t_1} g(k,l,S(t_1)) \| U(t_1) \| \| \Lambda_d^{1/2} \| \| \bar{V}^{\circ,\top}_{S}(t_1)   \widehat{v}^{\circ}_{l,S}(t_0)\|\\
\leq & \frac{1}{a}\left(\frac{c_1\lambda_1}{\lambda_l+1}\right)^{1/2}  \| \bar{V}^{\circ,\top}_{S}(t_1)   \widehat{v}^{\circ}_{l,S}(t_0)\| \beta_l(t_1)\frac{\log(n\vee t)}{\log(n\vee t_1)}.
\end{split}
\end{equation*} 
In the similar way  term $q_{22}$  may be  bounded  as follows
\begin{equation*}
\begin{split}
&q_{22}  \leq \left \|\frac{1}{t_1}\Xi_{k} \Xi_{S(t_1)}^{\top}    {V}^{\circ}_{S}(t_1)\right \| \| {V}^{\circ,\top}_{S}(t_1) \widehat{v}^{\circ}_{l,S}(t_0)  \| \\
&\leq  \frac{b}{a}  (1+o(1))\sqrt{\frac{c_2}{\lambda_l+1}}\beta_j(t_1)\frac{\log(n\vee t)}{\log(n\vee t_1)}\| {V}^{\circ,\top}_{S}(t_1) \widehat{v}^{\circ}_{l,S}(t_0) \|.
\end{split}
\end{equation*} 
The bound  on the term $q_{23}$ due to  \eqref{rowl12} reads as 
\begin{equation*}
\begin{split}
|q_{23}| = 
& \frac{1}{t_1}\| V_{k}\Lambda_d^{1/2}\|  \| U(t_1)^{\top}\Xi_{S}^{\top}(t_1)   {V}^{\circ}_{S}(t_1)\|\| {V}^{\circ,\top}_{S}(t_1)   \widehat{v}^{\circ}_{l,S}(t_0) \|\\
\leq & 2\sqrt{\frac{\log (n\vee t_1)}{t_1}}\| V_{k}\Lambda_d^{1/2}\|  \| {V}^{\circ,\top}_{S}(t_1)   \widehat{v}^{\circ}_{l,S}(t_0) \|\\
=& o(\beta_l(t_1)).
\end{split}
\end{equation*}  
Similarly to the case of $q_{14}$ one can show that 
\begin{equation*}
\begin{split}
|q_{24}| \leq & \|\frac{1}{t_1} \Xi_{k} U(t_1)\| \| \Lambda_d^{1/2}V_{S}^{\top}(t_1)    {V}^{\circ}_{S}(t_1) \| \| {V}^{\circ,\top}_{S}(t_1)   \widehat{v}^{\circ}_{l,S}(t_0)\| \\
\leq & \sqrt{\frac{c_2\lambda_1}{\lambda_l+1}} \frac{1}{a}\beta_l(t_1)\frac{\log(n\vee t)}{\log(n\vee t_1)} \| {V}^{\circ,\top}_{S}(t_1)   \widehat{v}^{\circ}_{l,S}(t_0)\|.
\end{split}
\end{equation*}  
Note that  (see \cite{ma2013sparse})    $ \| {V}^{\circ,\top}_{S}(t_1)   \widehat{v}^{\circ}_{l,S}(t_0)\| = 1+ o(1)$ and  $\| \bar{V}^{\circ,\top}_{S}(t_1)   \widehat{v}^{\circ}_{l,S}(t_0)\| = o(1)$, that is from above bounds  $\sum_{i=1}^4 |q_{1i}|=o\left(\sum_{i=1}^4 |q_{2i}|\right)
$. 
%
Therefore 
\begin{equation*}
\begin{split}
 \widehat{\Upsilon}(t_1)_{k,l} \leq& \frac{b}{a}\beta_{l}(t_1)  \sqrt{\sum_{j=1}^d \frac{\lambda_j+1}{\lambda_l+1}} \\
 &+ \frac{1}{a}\beta_{l}(t_1) \sqrt{2c_1} \frac{\log(n\vee t)}{\log(n\vee t_1)} \sqrt{\frac{\lambda_1+1}{\lambda_l+1}} .
 \end{split}
 \end{equation*}
Observe that $\sqrt{\sum_{j=1}^d \frac{\lambda_j+1}{\lambda_l+1}}\leq \sqrt{\tau} \sqrt{d}$ and $\lambda_1/\lambda_j \leq \tau $.
 Let us bound $\log(n\vee t)$  by  $\log(n\vee T)$ thus 
 \begin{align*}
 a \geq   \sqrt{2c_1}\frac{\log(n\vee T )}{\log (n\vee t_0)},\quad 
 b =   \frac{0.9 a- \sqrt{2c_1}\frac{\log(n\vee T )}{\log (n\vee t_0)}}{\sqrt{\tau} \sqrt{d}}.
 \end{align*}
 Therefore one gets for all $k\in N(t_1)$
 $$
 | \widehat{\Upsilon}(t_1)_{k,l}|\leq \beta_l(t_1)
 $$
 and $ \widehat{V}^{\circ}_N(t_0+1)=0$, and so, 
 $
 \widehat{V}^{\circ}_N(t_0+1)=\widehat{V}_N(t_0+1).
 $  
 
 To show  that 
 $$
\widehat{V}^{\circ}_N(u)=\widehat{V}_N(u),\quad   u=t_0+2,\dots,t 
 $$ 
we consider the events defined for the standard normal random variables  $z(k,l,S(u)) $ and $g(k,l,S(u))$ 
$$
\{ z(k,l,S(u)) \leq  \sqrt{c_1 \log (n \vee t)} \},
$$ 
 $$
\{ |g(k,l,S(u))| \leq \sqrt{ c_1 \log (n \vee t)}\}.
$$
Using the union bound 
\begin{equation*}
\begin{split}
&\mathbf{P}\left \{ \bigcup\limits_{\substack{l \in N(u), k=1,\dots,d,\\ u=t_0+1,\dots, t}} \left\{z(k,l,S(u))\leq  \sqrt{c_1 \log (n \vee t)}\right \} \right\} \\
&\leq  1-\sum_{k=1}^d\sum_{u=t_{0}+1}^{t}\sum_{l\in N(t)}\mathbf{P} \{z(k,l,S(u))  \leq \sqrt{c_1 \log (n \vee t)} \} \\ 
&\leq  1-n d (t-t_0) \mathbf{P} \{z(k,l,S(t))   \leq \sqrt{c_1 \log (n \vee t) } \}\\
&\leq 1-C_0 n (t-t_0) {\log(n\vee t)}^{-1}  { (n\vee t)}^{-c_1/2}. 
\end{split}
\end{equation*}
Take $c_1\geq 9$ to obtain the statement of the lemma.

\section*{Appendix B. Concentration of the spectral norm of the perturbation} 
Denote $\delta_{n,t}=\log(n\vee t)$.

\begin{lemma}  \cite{vershynin2010introduction} 
\label{norm_1}
Let  $X$ be a $t\times n$ matrix with i.i.d. $\mathcal{N}(0,1)$ entries. 
The following result holds true
 
\[
\mathbf{P}\left(\left\Vert\frac{1}{t}X^{\top}X-I_n \right\Vert \geq E_1(t,n,p)\right)\leq  2 (n\vee t)^{-p^2/2},
\]
where  
$$E_1(t,n,p)= 3\max\left(\sqrt{\frac{n}{t}}+p\frac{\sqrt{\delta_{n,t}}}{\sqrt{t}},\left[\sqrt{\frac{n}{t}}+\frac{p\sqrt{\delta_{n,t}}}{\sqrt{t}}\right]^{2}\right).$$
\end{lemma}

\begin{lemma} \cite{davidson2001local}
\label{norm_2}
Let $X$ and $Y$ be $t\times q$ and $t\times m$   matrices, $q>m$,  with i.i.d. $\mathcal{N}(0,1)$ entries then for any $0<x<1/2$ and $c>0$
\begin{align*}
\mathbf{P} \left(\|X^{\top}Y\| \geq t E_2(t,q,m,x,c) \right) \leq {\rm e}^{-\frac{c^2\delta_{n,t} }{2}}+ q {\rm e }^{-\frac{3x^2\delta_{n,t}}{16} } ,
\end{align*}
where 
$$E_2(t,q,m,x,c) =  \sqrt {1+x\frac{\delta_{n,t}}{t}} \left(\sqrt{\frac{q}{t}}+\sqrt{\frac{m}{t}}+c\frac{\sqrt{\delta_{n,t}}}{\sqrt{t}}\right).$$
\end{lemma}

\begin{lemma} \cite{ma2013sparse}
\label{sparse78}
There exist constants $\tilde{C}_1$ and $\bar{C}_1$ depending on $r$  and $\bar{C}_2$ and $\tilde{C}_2$ such that
\begin{align*}
E_1(t,{\rm card}(S(t)),p) &\leq \tilde{C}_1 \frac{\lambda_1M^{1/2}(t)}{\sqrt{t}}h_d+\tilde{C}_2  \sqrt{\frac{\delta_{n,t}}{t}}  
\\
E_2(t,{\rm card}({S(t)}),d,4,p) &\leq \bar{C}_1 \frac{ M^{1/2}(t)}{\sqrt{t}}\frac{h_d}{h_1}+\bar{C}_2  \sqrt{\frac{\delta_{n,t}}{t}} 
.
\end{align*} 
\end{lemma} 

\begin{theorem}(Wedin $\sin \theta$)
 \label{Wedin}  
  Let $A$ and $B$ be  $n\times k$, $n\geq k$, full-column rank matrices. Let the columns of a $n\times(n- k+1)$ matrix $U$ be the orthogonal matrices spanning the orthogonal complement of range of $B$. If the $\lambda_{\min}(A)\geq \epsilon \geq 0$ then 
  $$
   l(A,B) \leq \frac{\|A^{\top}U\|^2}{\epsilon^2}\leq \frac{\|B-A\|^2}{\epsilon^2}.
  $$
 \end{theorem}
\begin{theorem}(Davis $\sin\theta$)
\label{Davis}  \cite{davis1970rotation}
 Let $A$ and $B$ be the symmetric matrices with the decomposition $A  =W_{1} \Lambda_1 W_{1}^{\top}+ W_{2} \Lambda_2 W_{2}^{\top}$ and $B  =U_{1} \Delta_1 U_{1}^{\top}+ U_{2} \Delta_2 U_{2}^{\top}$, with conditions $[U_1, U_2]$ is orthogonal, $W_2$ is orthonormal and $W_1^{\top} W_2 = 0$,  the eigenvalues of $\Lambda_1W_1^{\top}W_1$ are contained in the interval $(a_1,a_2)$  and the eigenvalues of $\Delta_1$ are laying outside of the interval $(a_1-\epsilon,a_2-\epsilon)$ for some $\epsilon>0$ then 
 $$
 l(W_1,U_1) \leq \frac{\|U_2^{\top}(B-A)W_1\|^2}{\epsilon\lambda_{\min}^2(W_1)}.
 $$
\end{theorem}

\begin{lemma} 
\label{A1}
The norms of the subvectors $v_{j,N(t)}$ of $v_j$ satisfy 
\begin{equation*}
\|v_{j,N(t)}\|^2 \leq \left[\frac{2}{2-r}\frac{s_j^r}{(bh_j(t))^{r}}    \left(\frac{\delta_{n,t}}{t}\right)^{-r/2}  \wedge n \right] b^2 h_j^2 (t) \frac{\delta_{n,t}}{t}.
\end{equation*}
\end{lemma}

\begin{lemma}
Bound on the effective dimension ${\rm card} (S(t))$ is given by 
$$
d \leq {\rm card}(S(t))\leq C M(t)=C \left[n\wedge \sum_{j=1}^d  {s_j^r}{h_j^{-r} }\left( \frac{\delta_{n,t}}{t}\right)^{-r/2} \right].
$$
 \label{eff_dim}
\end{lemma}

\begin{lemma}
For a $\chi^2_{t}$ random variable $\zeta_t$  the following bounds hold \cite{johnstone2001chi}
\begin{align*}
\mathbf{P}(\zeta_t>t(1+\varepsilon)) &\leq {\rm e}^{-3t\varepsilon^2/16},\quad 0<\varepsilon<1/2,\\
\mathbf{P}(\zeta_t<t(1-\varepsilon)) &\leq {\rm e}^{- t\varepsilon^2/4},\quad 0<\varepsilon<1,\\
\mathbf{P}(\zeta_t>t(1+\varepsilon)) &\leq \frac{\sqrt{2}}{\varepsilon\sqrt{t}}{\rm e}^{- t\varepsilon^2/4},\quad 0<\varepsilon<t^{1/16},\,t>16.
\end{align*}
\end{lemma}

\section*{Acknowledgment}
This work is funded by the German Research Foundation (DFG), Collaborative Research Center 823, Subprojects B3 and C5.




%
\bibliographystyle{IEEEtran}
\bibliography{IEEEabrv,refs}

\end{document}